\pdfoutput=1
\documentclass{aa}
\usepackage{txfonts}

\usepackage{amssymb,epsfig,color}
\usepackage{amsmath}
\usepackage{graphicx}
\graphicspath{{figures/}{figures/paper/}}

\usepackage{makecell}

\usepackage{siunitx}
\usepackage{booktabs}
\usepackage{multirow}
\DeclareSIUnit\arcsecond{arcsec}
\DeclareSIUnit\parsec{pc}
\DeclareSIUnit\milliarcsecond{mas}
\DeclareSIUnit\year{yr}
\usepackage{longtable}
\usepackage{lscape}

 \usepackage{flushend}

\usepackage[hyperindex,breaklinks=true, colorlinks,citecolor=blue,draft=False]{hyperref}

\usepackage{textcomp}
\newcommand{\micron}{\textmu m}

\usepackage{xcolor}

\def\reff@jnl#1{{\rm#1\/}}

\def\aj{\reff@jnl{AJ}}
\def\araa{\reff@jnl{ARA\&A}}
\def\apj{\reff@jnl{ApJ}}
\def\apjl{\reff@jnl{ApJ}}
\def\apjs{\reff@jnl{ApJS}}
\def\ao{\reff@jnl{Appl.Optics}}
\def\apss{\reff@jnl{Ap\&SS}}
\def\aap{\reff@jnl{A\&A}}
\def\aapr{\reff@jnl{A\&A\simRev.}}
\def\aaps{\reff@jnl{A\&AS}}
\def\azh{\reff@jnl{AZh}}
\def\baas{\reff@jnl{BAAS}}
\def\jrasc{\reff@jnl{JRASC}}
\def\memras{\reff@jnl{MmRAS}}
\def\mnras{\reff@jnl{MNRAS}}
\def\nar{\reff@jnl{New Astronomy Reviews}}
\def\pra{\reff@jnl{Phys.Rev.A}}
\def\prb{\reff@jnl{Phys.Rev.B}}
\def\prc{\reff@jnl{Phys.Rev.C}}
\def\prd{\reff@jnl{Phys.Rev.D}}
\def\prl{\reff@jnl{Phys.Rev.Lett}}
\def\pasa{\reff@jnl{PASA}}
\def\pasp{\reff@jnl{PASP}}
\def\pasj{\reff@jnl{PASJ}}
\def\qjras{\reff@jnl{QJRAS}}
\def\skytel{\reff@jnl{S\&T}}
\def\solphys{\reff@jnl{Solar\simPhys.}}
\def\sovast{\reff@jnl{Soviet\simAst.}}
 \def\ssr{\reff@jnl{Space\simSci.Rev.}}
\def\zap{\reff@jnl{ZAp}}
\def\nat{\reff@jnl{Nature}}

\begin{document}

    \title{A Full-Sky SPHEREx Map of Integrated 3.3 and 3.4\,\micron\ PAH Emission}

    \author{
      Roke Cepeda-Arroita\inst{1,2}\thanks{\email{roke.cepeda@iac.es}}
          \and
      Clive Dickinson\inst{3}
      \and
      J. Alberto Rubi\~no-Mart\'{i}n\inst{1,2}
      \and
      Ricardo T. G\'{e}nova-Santos\inst{1,2}
      \and
      \newline
      Gabriel A. Hoerning\inst{3}
      \and
      Stuart E. Harper\inst{3}
      \and
      Debabrata Adak\inst{1,2}
    }

	\institute{
		Instituto de Astrof\'{i}sica de Canarias, 38200 La Laguna, Tenerife, Canary Islands, Spain
		\and
		Departamento de Astrof\'{i}sica, Universidad de La Laguna (ULL), 38206 La Laguna, Tenerife, Spain
		\and
		Jodrell Bank Centre for Astrophysics, Alan Turing Building, Department of Physics and Astronomy, The University of Manchester, Oxford Road, Manchester M13 9PL, UK
	}

	\date{Received XXX; accepted XXX}

\abstract{
Polycyclic aromatic hydrocarbons (PAHs) are a ubiquitous component of the interstellar medium and are among its smallest carbonaceous grains. Their infrared emission bands trace this small-grain population together with the interstellar radiation field that excites it, and are widely used as diagnostics of photodissociation regions, of dust processing and of star formation. We present full-sky maps of the integrated radiance of the 3.3~\micron\ aromatic C--H stretch and the adjacent 3.4--3.5~\micron\ aliphatic complex, constructed from SPHEREx QR2 Level-2 data. Existing all-sky PAH tracers are broadband, so the band emission can only be inferred as an excess over stellar and continuum emission; SPHEREx samples the spectrum finely enough for the bands to be fitted directly. We mask flagged detector pixels, subtract the per-pixel zodiacal-light model, remove point sources by fitting an empirically derived PSF, and coadd the results into inverse-variance-weighted HEALPix maps at $2.6\arcmin$ FWHM and $N_{\rm side}=4096$ in 102 narrowband channels across Detector~4 ($2.42$--$3.82$~\micron). At every sky pixel we fit a continuum and the two bands to the resulting spectrum. The combined band radiance is the primary product; its decomposition into the aromatic and aliphatic components is released alongside it as an estimate, with its modelling systematic stated. These are the first full-sky maps of PAH band emission, resolving the 3.3 and 3.4--3.5~\micron\ bands spectrally, and the first full-sky PAH maps from SPHEREx to be released. Because the 3.3~\micron\ band is emitted preferentially by the smallest PAHs, the maps trace directly a grain population, including the candidate carriers of anomalous microwave emission, that broadband PAH proxies reach only indirectly. Further improvements are expected as the survey accumulates coverage. The maps and their associated products, including a reliability mask, are publicly available with an explanatory supplement at \url{https://research.iac.es/proyecto/radioforegroundsplus/pages/data-products/spherex-pahs.php}.
}


\maketitle

\section{Introduction}
\label{sec:intro}

Polycyclic aromatic hydrocarbons (PAHs) are thought to carry the family of infrared emission features at 3.3, 6.2, 7.7, 8.6, 11.3 and 12.7~\micron\ that dominate the mid-infrared spectrum of almost any region of the interstellar medium exposed to ultraviolet light \citep{Leger1984,Allamandola1985,Tielens2008}. The emission is stochastic in origin: absorption of a single ultraviolet photon transiently heats a very small grain to of order $10^{3}$~K \citep{Sellgren1984}, and in the PAH interpretation that grain is a molecule of some tens to hundreds of carbon atoms, which re-radiates the energy in the vibrational modes of its carbon skeleton and of its peripheral C--H bonds \citep{Leger1984,Allamandola1985,Draine2007}. The bands therefore trace the smallest carbonaceous grains together with the radiation field that excites them, and are used to diagnose the PAH abundance, the intensity of the starlight that heats them and the local physical conditions \citep{Draine2007,Li2020}, and as a tracer of star formation \citep{Li2020}.

The 3.3~\micron\ aromatic C--H stretch is the shortest-wavelength member of the family and the most demanding in excitation temperature, so it is emitted preferentially by the smallest PAHs \citep{Tielens2008,Draine2007}. Like the other PAH bands, it is far less attenuated by interstellar extinction than optical and ultraviolet tracers, so it can be followed through the Galactic plane as well as at high latitude. A full-sky map of this band therefore traces the smallest carbonaceous grains directly: their abundance relative to the rest of the dust, their survival in ionised gas and their processing by the radiation field, which all-sky data have so far reached only through broadband photometry of the longer-wavelength bands, emitted mainly by larger PAHs \citep{Draine2007}. The 3.4--3.5~\micron\ complex immediately redward of it is attributed to aliphatic C--H stretching modes, whether of methyl side groups attached to the aromatic skeleton \citep{Joblin1996} or of extra hydrogen atoms added to it \citep{Bernstein1996}. Its strength relative to the 3.3~\micron\ band measures the aliphatic content of the grains, which depends on their environment and history: aliphatic groups are removed by ultraviolet photo-processing more readily than the aromatic skeleton, so the ratio falls towards the exciting stars of reflection nebulae \citep{Joblin1996} and from dense into diffuse gas \citep{Boersma2026}, whereas some carbon-rich post-AGB objects show a 3.4~\micron\ band comparable in flux to the 3.3~\micron\ one, and aliphatic C--H absorption is seen along diffuse lines of sight \citep{Chiar2013}. The two features are close enough that any instrument able to separate them measures both at once.

The smallest PAHs are also the candidate carriers of anomalous microwave emission (AME), an important Galactic foreground at 10--60~GHz generally attributed to electric dipole radiation from rapidly spinning ultrasmall grains \citep{Draine1998,Dickinson2018}. Whether they are its dominant carrier is unsettled. \citet{Hensley2016} found no correlation between the AME per unit dust radiance and the ratio of the WISE 12~\micron\ emission to the dust radiance, an indirect tracer of PAH abundance, whereas a survey of 144 Galactic clouds with S-PASS, C-BASS and QUIJOTE data between 2.3 and 20~GHz, which constrain the peak frequency and width of the AME spectrum, recovers correlations with similar tracers \citep{CepedaArroita2026}, and differences in PAH emission physics between environments could weaken a correlation even if PAHs do carry the AME \citep{Hensley2022}. The all-sky null result rests on a broadband tracer of larger PAHs and on Commander AME amplitudes \citep{Planck2015_X} that, with no data between $408$~MHz and $23$~GHz, are by that analysis' own assessment degenerate with free-free and synchrotron emission. A map of the band emitted by the smallest PAHs, measured spectrally, is the observable a decisive test requires.

At near- and mid-infrared wavelengths, all-sky coverage has so far been photometric rather than spectroscopic: COBE/DIRBE mapped the sky in ten bands from 1.25 to 240~\micron\ \citep{Hauser1998}, WISE at 3.4--22~\micron\ \citep{Wright2010}, and IRAS \citep{Neugebauer1984} and AKARI \citep{Murakami2007} at mid- and far-infrared wavelengths. The spectroscopy that does exist over large areas either is restricted to point sources or lies at longer wavelengths: the IRAS Low Resolution Spectrometer recorded 7.7--22.6~\micron\ spectra of the brighter point sources \citep{Olnon1986}, and COBE/FIRAS measured the spectrum of the sky between 0.1~mm and 1~cm \citep{Mather1994}. The most widely used all-sky PAH proxy, and the one on which the test of \citet{Hensley2016} rests, is the WISE 12~\micron\ atlas of \citet{MeisnerFinkbeiner2014}, whose band contains the 11.3~\micron\ feature but also continuum emission, so that the PAH contribution has to be separated from the rest by modelling. Spectroscopic isolation of the bands themselves has come from pointed instruments over individual fields: ISO \citep{Kessler1996} and the AKARI Infrared Camera \citep{Onaka2007} for the 3.3~\micron\ band as well as the longer-wavelength ones, and Spitzer \citep{Werner2004}, whose spectrograph does not reach 3.3~\micron, for the mid-infrared bands; and from the IRTS spectrometers, which detected the 3.3~\micron\ band \citep{Tanaka1996} and the mid-infrared bands \citep{Onaka1996} in diffuse Galactic emission along their scan path, but over a small fraction of the sky.

SPHEREx \citep{Crill2020,Bock2026} removes that constraint. Because it disperses in position rather than in time, it returns a low-resolution spectrum from $0.75$ to $5.0$~\micron\ at every point on the sky rather than a set of broadband fluxes, and it does so over the whole sky roughly twice a year. The Level-1 and Level-2 processing that turns the raw frames into calibrated, astrometrically and spectrally solved images is described by \citet{Akeson2026}, and the mission's own approach to building sky maps from them by \citet{Cukierman2026}.

\citet{Zhang2025} forecast what the survey should recover of the 3.3~\micron\ band in nearby galaxies. The first SPHEREx map of diffuse 3.3~\micron\ emission on Galactic scales is that of \citet{Murgia2026}, which established what the instrument can do for diffuse emission: a strong correlation of the band with the Planck thermal dust radiance, a companion map of Brackett-$\alpha$ at $4.05$~\micron, and a PAH abundance map showing systematic depletion of PAHs within ionised gas across the Galactic plane. Their analysis covers the $30\%$ of the sky around the Galactic plane that the Planck 70\% Galactic mask excludes, at $N_{\rm side}=512$ and $20\arcmin$ resolution, and the band is measured as the intensity interpolated to $3.3$~\micron\ minus a linear continuum through maps interpolated to $3.15$ and $3.60$~\micron; the authors present the result as an initial step towards full-sky diffuse-emission maps. Slightly earlier SPHEREx studies had mapped the band over individual regions: \citet{Boersma2026} separate its aromatic and aliphatic components across the north-western photodissociation region of the Iris Nebula, and \citet{Hora2026} map ices and PAHs over the Cygnus~X and North America Nebula fields. \citet{Li2026} has since mapped the 3.3~\micron\ feature across the Magellanic Clouds.

This paper takes the next step, and differs from \citet{Murgia2026} in three respects. It covers the whole sky, as imaged by SPHEREx up to July 2026, rather than the region around the Galactic plane. It is built at $2.6\arcmin$ FWHM on an $N_{\rm side}=4096$ grid, close to the finest resolution that pixelisation supports and a factor of about eight finer in beam. And the band is measured by fitting, at every sky pixel, a continuum together with a model of the two features, an instrument-convolved Drude profile for the 3.3~\micron\ band and a Gaussian for the 3.4--3.5~\micron\ complex, to a spectrum sampled by 102 narrowband channels across Detector~4 alone, rather than by differencing the intensity at the band centre against a two-point continuum. That last difference is not only one of precision. At the nominal resolving power of $R=35$ the 3.3~\micron\ feature is narrower than one resolution element, so what a single-wavelength estimate returns depends on where the continuum is placed; and an intensity evaluated at 3.3~\micron\ does not measure the aliphatic complex separately, while a continuum anchored at 3.6~\micron\ sits inside it (Sect.~\ref{sec:methods-pah-fitting}). Fitting the sampled line profile instead returns the two components separately, with the instrumental width carried explicitly, and their sum as the integrated radiance of the 3~\micron\ complex.

The map-making shares its architecture with \citet{Cukierman2026}. Each tile is processed in its own detector frame before any reprojection; the wavelength axis is defined by logarithmically spaced channels evaluated on the per-pixel spectral solution; and the sky is built up by accumulating tile pixels additively onto a HEALPix grid \citep{Gorski2005}, so that overlapping exposures combine without special treatment and the result does not depend on the order in which tiles are processed. The weighting differs: \citet{Cukierman2026} coadd with an intensity-weighted histogram and a hit count, whereas we accumulate inverse-variance weighted sums and recover the variance of the mean alongside the map. It differs in three choices that follow from targeting a narrow feature at high angular resolution. The channelisation is fixed at 102 bins across Detector~4 alone, about six times finer than the instrument resolves, rather than the nominal 17 per detector, which \citet{Cukierman2026} subdivide six-fold in their example of a narrow emission line, which samples the instrumental line profile and limits the spread of effective wavelength contributing to a channel (Sect.~\ref{sec:data-band4}). Point sources are fitted and subtracted with an empirical PSF rather than masked, so that the diffuse sky under and around bright stars is retained rather than removed with them (Sect.~\ref{sec:methods-starremoval}). And the zodiacal light is treated spectrally before it is treated spatially: the Level-2 zodiacal model is subtracted per pixel, what survives it is spectrally smooth across Detector~4 and is partially absorbed by the continuum term of the per-pixel fit, and only the degree-scale stripes this leaves in faint sky are filtered from the coadded maps, whose unfiltered versions are released alongside (Sects.~\ref{sec:methods-zodi}, \ref{sec:methods-pah} and \ref{sec:postproc-destripe}).

The cost of working at this resolution and this spectral sampling is volume. The observatory integrates in $116.9$~s exposures \citep{Bock2026}, each of which produces one Level-2 tile per detector of $72$~MB \citep{Akeson2026}, and the spectral images are the largest public SPHEREx product, totalling $190$~TB over the 25-month prime mission \citep{Akeson2026}, corresponding to $\sim4.4\times10^{5}$ exposures, or about $78\%$ of what uninterrupted $116.9$~s integrations would give over that period, the remainder presumably going to operational overheads such as slews between pointings. QR2 is a partial release: the Detector-4 tiles used here amount to about $16$~TB, or $226\,620$ tiles (one per exposure), out of the current $\sim100$~TB across all six detectors, that is, close to half of the eventual volume. Every step described below is therefore applied per tile and folded immediately into an additive, resumable accumulator, which is what makes a full-sky run at $N_{\rm side}=4096$ tractable on modest hardware and splittable across machines.

The rest of the paper is organised as follows. Section~\ref{sec:data} describes the SPHEREx survey, the Level-2 data model, the choice of Detector~4 and its narrowband channelisation, and the ancillary catalogues. Section~\ref{sec:methods} sets out the processing: per-tile masking and zodiacal-light subtraction, stellar subtraction with an empirical PSF, regridding and full-sky accumulation, the per-pixel spectral extraction of the two features, and the post-processing applied to the coadded maps. Section~\ref{sec:results} presents the full-sky map, its residual systematics, the comparison with WISE and the reliability mask released with the maps. Section~\ref{sec:conclusions} summarises the results and the limitations carried forward.

\section{Data}
\label{sec:data}

\subsection{The SPHEREx All-Sky Spectral Survey}
\label{sec:data-survey}

SPHEREx (Spectro-Photometer for the History of the Universe, Epoch of Reionization and Ices Explorer) is a NASA Medium Explorer launched on 11 March 2025 \citep{Bock2026}, which is carrying out the first all-sky spectral survey in the near infrared, covering $0.75$--$5.0$~\micron\ \citep{Crill2020,Bock2026}. The telescope has a $20$~cm aperture and images the sky onto six detector arrays, each behind its own linear variable filter (LVF), with a native pixel scale of $6.15\arcsec$. Because an LVF disperses in position rather than in time, the central wavelength varies across every tile, and a given sky position is observed at one wavelength per exposure and at many wavelengths over the course of the survey. The spacecraft is in a Sun-synchronous polar orbit, so a single orbit sweeps from pole to pole while the orbital plane precesses in ecliptic longitude at approximately the solar rate. The whole sky is covered roughly twice per year, giving four full-sky surveys over the nominal two-year mission \citep{Bock2026}.

We use the publicly released Quick Release 2 (QR2) Level-2 products: calibrated, astrometrically solved spectral images produced by the SPHEREx Science Data Center and distributed by IRSA, as available on September 2026. Level-2 spectral images reach the archive within about 60 days of the observations \citep{Akeson2026}, so the holdings used here extend to approximately July 2026. QR2 covers only part of the nominal mission, so the maps presented here are built from partial coverage, with the consequences discussed in Sects.~\ref{sec:methods-healpix-accum} and \ref{sec:results}.

\subsection{Level-2 Spectral Images}
\label{sec:data-tiles}

Each Level-2 spectral image, hereafter a tile, is a multi-extension FITS file of $2040\times2040$ pixels covering approximately $3.5\degr\times3.5\degr$. It carries the calibrated surface brightness in MJy\,sr$^{-1}$, a per-pixel flag bitmask, a per-pixel variance, a model of the zodiacal light in the same units which is not pre-subtracted, an oversampled PSF cube, and a lookup table from which the spectral solution is evaluated.

\begin{figure*}[t]
  \centering
  \includegraphics[width=\textwidth]{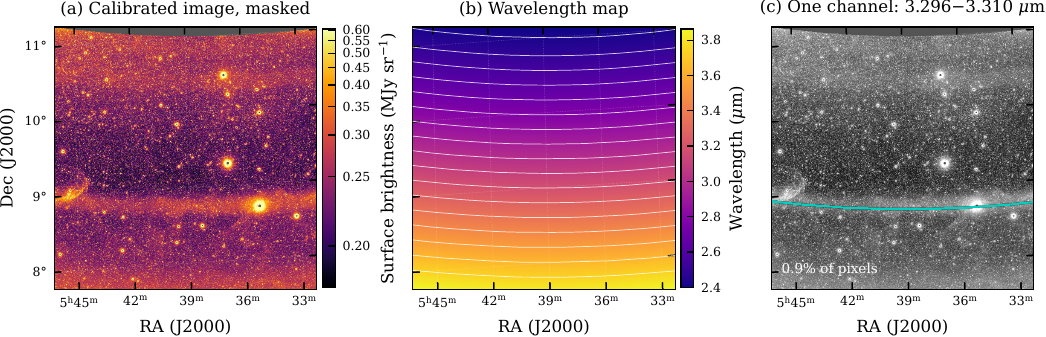}
  \caption{One SPHEREx Detector-4 Level-2 tile of the Barnard~30 / $\lambda$~Orionis field (observation 2025W37\_1A\_0276\_1). \textbf{(a)} Calibrated surface brightness, with the flagged pixels masked and shown in grey. \textbf{(b)} The per-pixel wavelength solution of the linear variable filter, with every sixth narrowband channel boundary drawn in white. \textbf{(c)} The pixels of the same tile that fall within a single narrowband channel, marked in colour over the greyscale image. Each channel samples a narrow strip of any one exposure, here $0.9\%$ of the array, so full-sky coverage of a given channel accumulates only as the survey revisits each sky position with the filter at different positions.}
  \label{fig:tile-overview}
\end{figure*}

Two coordinate solutions apply to the same array. The spatial one is a standard tangent-plane WCS with polynomial distortion terms, giving the sky position of each pixel. The spectral one gives the central wavelength and the bandwidth of each pixel. This second solution is what makes the present work possible: the wavelength is set by the position on the detector, not by the exposure, and each exposure places a given sky position at a different place on the detector, so combining many exposures of that position builds up a sampled spectrum. Figure~\ref{fig:tile-overview} shows one such tile together with its wavelength solution.

Tiles are located and retrieved through the IRSA simple image access (SIA) service\footnote{\url{https://irsa.ipac.caltech.edu/applications/spherex/}}, with one query per position on the sky grid described in Sect.~\ref{sec:methods-healpix-accum}.

\subsection{Detector 4 and the Narrowband Channels}
\label{sec:data-band4}

We refer to the six arrays as detectors, numbered 1 to 6, and reserve the word channel for the narrow wavelength bins into which we divide a detector's range. The detectors cover $0.75$--$1.11$, $1.11$--$1.64$, $1.64$--$2.42$, $2.42$--$3.82$, $3.82$--$4.42$ and $4.42$--$5.00$~\micron, with resolving powers of $R=41$, 41, 41, 35, 110 and 130 respectively \citep{Bock2026}.

All of the work presented here uses Detector~4, which contains both the 3.3~\micron\ aromatic C--H stretch and the 3.4--3.5~\micron\ aliphatic complex within a single continuous LVF. Keeping both features on one detector avoids having to stitch together two independently calibrated bands across the dichroic boundary.

The nominal spectral sampling of the survey is 17 channels per detector, or 102 across all six, and \citet{Cukierman2026} divide each nominal channel into six, for 102 per detector, in their example of a narrow emission line. We adopt the finer sampling throughout, dividing Detector~4 into 102 channels, spaced logarithmically so that $\Delta\ln\lambda$ is constant and the channel resolving power $\lambda/\Delta\lambda\approx220$ is the same across the band. These channels are deliberately narrower than the spectral resolution element of the instrument, which at $R=35$ is about six times wider than one channel. Sampling this finely adds no spectral information that the LVF has not already recorded, but it does two things that matter here. It samples the instrumental line profile finely enough for the shape of the features to be fitted rather than merely integrated (Sect.~\ref{sec:methods-pah}), and it reduces the spread of wavelengths contributing to a channel, which would otherwise imprint the LVF gradient on the maps as a position-dependent effective wavelength. The cost is coverage, and computational cost: a finer channelisation multiplies the number of accumulator planes that must be held and written. A narrower channel admits fewer detector pixels per exposure, as Fig.~\ref{fig:tile-overview}c illustrates, so each channel fills in more slowly than a nominal one, which is the dominant limitation at QR2 depth (Sect.~\ref{sec:methods-healpix-accum}).

\subsection{Ancillary Catalogues}
\label{sec:data-ancillary}

Stellar subtraction (Sect.~\ref{sec:methods-starremoval}) uses the AllWISE Source Catalogue \citep{Wright2010,Cutri2013} as its input source list, through the $W1$ magnitude at $3.4$~\micron, the WISE band closest in wavelength to Detector~4. The small number of stars too bright for AllWISE photometry is supplemented with the Yale Bright Star Catalogue \citep{Hoffleit1991}, as described in Sect.~\ref{sec:methods-ultrabright}. WISE also provides the independent broadband comparison map used in Sect.~\ref{sec:discussion-wise}.

\section{Methods}
\label{sec:methods}

\subsection{Per-Tile Pre-Processing}
\label{sec:methods-preprocessing}

The pipeline runs in two stages. The first turns raw Level-2 tiles into full-sky, per-channel, inverse-variance-weighted HEALPix maps (Sects.~\ref{sec:methods-preprocessing} to \ref{sec:methods-healpix}); the second performs the per-pixel spectral extraction on those maps (Sect.~\ref{sec:methods-pah}). Everything in the first stage is done tile by tile, in the frame in which each tile was taken, so that the PSF and the LVF wavelength solution are only ever used where they are defined. Reprojection onto the sky grid is the last operation applied to a tile.

\subsubsection{Pixel Masking}
\label{sec:methods-flags}

The flag extension records per-pixel conditions as a bitmask. We treat as invalid the pixels marked as saturated or in detector-ramp overflow, affected by on-board processing errors, permanently non-functional, hot or cold, reference pixels, of low efficiency near the dichroic edge, missing, with a failed nonlinearity correction, carrying persistent charge above threshold, hit by transients, identified as phantom pixels, or flagged as outliers. The bits themselves are defined by the SPHEREx Science Data Center; the pipeline paper \citep{Akeson2026} describes the principal ones and the complete list is given in the SPHEREx Explanatory Supplement \citep{SPHERExExpSupp2025}, so we do not reproduce them here. Together they invalidate $4.50\%$ of the array on a representative Detector-4 tile, dominated by the strip of low efficiency along the dichroic edge, which alone accounts for $2.70\%$ and is a fixed region of the array rather than a property of the observation. One bit is deliberately not treated as invalid: the flag marking pixels mapped to a catalogued source covers $80.87\%$ of that tile, and masking it would leave the median filter of Sect.~\ref{sec:methods-median} with gaps exactly where the stars it is meant to suppress are. Invalid pixels are set to NaN in both the tile and the variance, and every subsequent operation is NaN-aware, so that a masked pixel never contributes to a fit, a filter or a coadd.

\subsubsection{Zodiacal-Light Subtraction}
\label{sec:methods-zodi}

Zodiacal light is the dominant diffuse foreground over most of the sky at these wavelengths. The Level-2 products carry a per-pixel model of it, produced by the SPHEREx Science Data Center from an empirical model of the zodiacal cloud rather than from the SPHEREx data themselves \citep{SPHERExExpSupp2025}; it is not pre-subtracted, and we subtract it directly. The model is treated as noiseless, so the variance is unchanged by this step.

Subtracting the template at unit amplitude is the simplest available option and it is not exact. One expected source of mismatch is the seasonal variation of the zodiacal cloud, which the model does not fully capture (A. Cukierman, private communication), so that the residual is not a static pattern on the sky but depends on when a given line of sight was observed. Wherever the model and the real sky disagree, the difference stays in the tile, and away from the Galactic plane, where the diffuse PAH signal is weakest, that residual can exceed the signal in an individual channel. Two considerations make this acceptable here. The first is spectral: zodiacal light is a combination of scattered sunlight and thermal emission from interplanetary dust, both of which are smooth across the range of Detector~4, with no structure on the scale of the line window. A mismatch in the model therefore appears in a pixel's spectrum as a smooth offset or slope, which the continuum fit of Sect.~\ref{sec:methods-pah-fitting} absorbs, rather than as a feature that could be mistaken for band emission. The second is that the spatial pattern the residual leaves, degree-scale stripes along the ecliptic parallels, is removed afterwards from the coadded maps by a filter calibrated to conserve the sky (Sect.~\ref{sec:postproc-destripe}), and the unfiltered maps are released alongside the filtered ones. Fitting the amplitude of the model and a monopole offset against the data, rather than subtracting it, is the natural refinement for a future release.

\subsection{Stellar Source Removal}
\label{sec:methods-starremoval}

Across Detector~4 the sky is dominated by stars, and increasingly so towards the blue: the stellar contribution rises steeply and the diffuse contribution falls across the SPHEREx wavelength range, so the problem described here is milder on Detector~4 than it would be on the shorter-wavelength detectors. Even at high Galactic latitude, a single SPHEREx tile contains thousands of catalogued point sources whose wings extend over tens of pixels, and the diffuse emission we are after is orders of magnitude fainter. Masking the sources is not sufficient: the brightest stars would take a substantial fraction of the sky with them, and the extended wings that dominate the contamination of the surrounding diffuse background would survive any mask small enough to be acceptable. We therefore fit and subtract an empirical point-spread function (PSF) at the position of every catalogued source above a brightness limit, and treat what is left below that limit statistically. This is done per tile, before any reprojection, so that the subtraction is performed in the frame in which the PSF is actually defined.

\subsubsection{Empirical PSF Construction}
\label{sec:methods-psf-construction}

\begin{figure*}[t]
  \centering
  \includegraphics[width=0.9\textwidth]{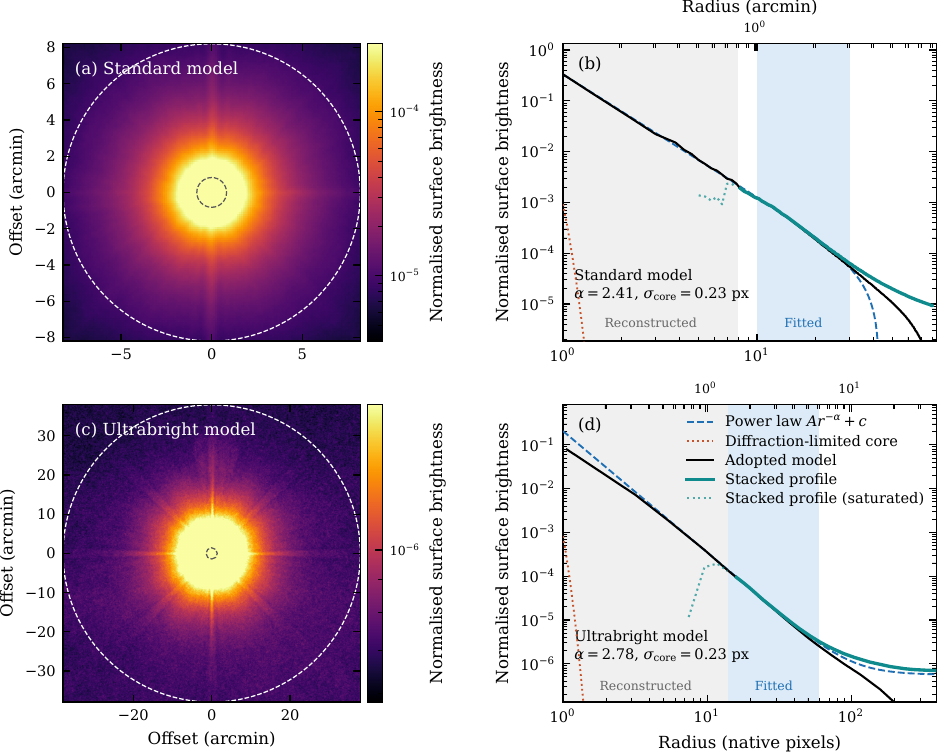}
  \caption{Empirical PSF models for Detector~4 and their reconstructed cores. \textbf{(a)} and \textbf{(c)}: the standard and ultrabright stacks on a logarithmic stretch, with the core reconstructed but before the pedestal subtraction. The dashed grey circle marks the radius inside which the stack is saturated and the model is reconstructed, and the dashed white circle the radius beyond which the adopted model is set to zero. The colour scales are logarithmic and run from $4.0\times10^{-6}$ to $2.6\times10^{-4}$ in \textbf{(a)} and from $5.8\times10^{-7}$ to $1.7\times10^{-6}$ in \textbf{(c)}, in units of the unit-normalised stack, with only whole decades labelled. \textbf{(b)} and \textbf{(d)}: azimuthal median profile of the stack, the power law fitted over the shaded annulus, the diffraction-limited Gaussian core of fixed width, and the adopted model. The reconstructed region is shaded in grey. Note the different radial ranges of the two models. The small departure of the adopted model from the fitted power law inside $r\approx4$~pixels is of no consequence: the cores of all the contributing stars are saturated and masked well outside that radius, so no measurement exists there to be reproduced.}
  \label{fig:psf-model}
\end{figure*}

Subtracting a star at the level required here means modelling it from the saturated core out to wings that extend some two orders of magnitude further in radius than the instrumental FWHM would naively suggest. The oversampled instrumental PSF distributed with each Level-2 tile cannot do this on its own: once binned to the native $6.15\arcsec$ pixels and normalised to match the observed wings, its usable extent is only about $10\times10$ pixels, or $1.0\arcmin\times1.0\arcmin$. That is far smaller than the radius over which a bright star actually deposits light, so the distributed PSF is adequate only for the faintest sources and cannot be used for the stars that dominate the contamination. We therefore construct the model empirically and reconstruct only its unresolved core from theory. Figure~\ref{fig:psf-model} shows the two models built this way and the profiles behind them.

The empirical model is built by stacking bright stars at high Galactic latitude, where the diffuse background under each star is faint and smooth. The stacked stars span $-1.81 < W1 < -0.47$ in the WISE $3.4$~\micron\ magnitude $W1$, and $15\degr < |b| < 89\degr$. Following the bright-source selection of \citet{MeisnerFinkbeiner2014}, who applied the same magnitude cuts in the WISE 12~\micron\ band ($W3$) for their atlas to select roughly the 100th--1100th brightest high-latitude sources, we draw AllWISE sources \citep{Wright2010,Cutri2013} with $-2 < W1 < 1$ and $|b| > 15\degr$, rank them by increasing $W1$, and stack those ranked 101--1101; the hundred brightest are excluded here and handled separately in Sect.~\ref{sec:methods-ultrabright}. For each source we extract a $161\times161$~pixel ($\pm8.2\arcmin$) cutout from a Detector-4 tile to which the invalid-pixel mask and zodiacal-light subtraction of Sect.~\ref{sec:methods-preprocessing} have already been applied, recentre it to sub-pixel accuracy, and normalise it to unit integrated flux. Before stacking, each cutout is rescaled radially by $\lambda_{\rm mid}/\lambda$, where $\lambda$ is the wavelength observed at the star's detector position and $\lambda_{\rm mid}=3.12$~\micron\ is the band centre. This step is specific to an LVF instrument: wavelength maps onto position along the dispersion axis, the PSF width scales with wavelength, and a stack of stars observed at different detector positions would otherwise be a blend of PSFs of different sizes. The rescaled cutouts are combined with a pixel-wise median, which rejects the companions, the residual structure present in any individual stamp, and any transient not already caught by the Level-2 flag mask.

The radial scaling is a first-order correction rather than an exact one. A purely diffractive response scales linearly with wavelength, and tests on the stack confirm that this approximation holds well over Detector~4. Not every component of the response does so, however: internal reflections in particular need not scale linearly, and can depend on where the star sits relative to the centre of the array rather than only on its wavelength. The true wavelength dependence is therefore warped rather than a pure dilation, and we correct only its leading term. This is one reason the ultrabright stars, whose models are dominated by exactly these non-diffractive components, are the hardest to reproduce. A stack built on the exact per-pixel wavelength solution and the measured LVF response, rather than on a single radial rescaling, is left as future work.

Within $r\approx8$~pixels the stack is not trustworthy, since every contributing star is saturated there and those pixels are already flagged and masked in the Level-2 products, so the stack carries no information at all inside that radius. Over the surrounding annulus the measured profile follows a power law, and we fit $I(r)=A\,r^{-\alpha}+c$ to the azimuthally averaged profile between 10 and 30~pixels, obtaining $\alpha=2.41$. Over that annulus the law reproduces the measured profile with a fractional scatter of $5.8\%$ rms, and the corresponding number for the ultrabright stack of Sect.~\ref{sec:methods-ultrabright}, fitted between 14 and 60~pixels, is $2.6\%$. The index itself is less well determined than these residuals suggest: refitting the standard stack over $10$--$25$ or $15$--$30$~pixels returns $\alpha=2.29$ and $\alpha=3.02$ respectively, so $\alpha$ is uncertain at the level of $\pm0.3$ through the choice of annulus alone, whereas the ultrabright index is stable to $\pm0.07$ against the same test. This matters less than it appears: extending the power law inward across the saturated region is an empirical statement about how steeply the response rises with decreasing radius, not a physical model of it, and that steepness is the one property the subtraction needs, obtained without a free parameter. A pure power law cannot represent the unresolved centre, so in both the standard model and the ultrabright one of Sect.~\ref{sec:methods-ultrabright}, below a handoff radius $R_{\rm core}$ the power law is replaced by a Gaussian whose width is fixed by diffraction rather than fitted: at the band centre the Airy core has ${\rm FWHM}=1.028\lambda/D=3.3\arcsec$ for the $D=0.2$~m aperture, i.e.\ $0.54$ native pixels, or $\sigma=0.23$~pixels. Requiring the value and the first derivative to be continuous at the join fixes $R_{\rm core}=\sigma\sqrt{\alpha}$ and the Gaussian amplitude, so no blending zone or free parameter is introduced. The exactness of the reconstruction in the innermost pixels is in any case not critical: no star in the stack, and no star bright enough to need this model, has unsaturated data inside $r\approx8$~pixels, so the reconstructed core is never compared against a measurement and is only ever used where the data are masked. Finally, the azimuthal median measured in a thin annulus at the outer radius of each model, $r=80$~pixels for the standard one and $370$~pixels for the ultrabright one, is subtracted from the whole model and the model is truncated to zero beyond that radius, so that the wings go to zero without an arbitrary taper distorting the profile inward of it. For the standard model the subtracted pedestal, $9.3\times10^{-6}$ of the unit-normalised stack, is comparable to the measured wing at $r=80$~pixels, $9.4\times10^{-6}$, so it removes part of the genuine wing along with any residual background. It is nonetheless needed: a model whose wings do not go to zero forces the amplitude fit of Sect.~\ref{sec:methods-psf-fitting} to absorb the offset, over-subtracting the diffuse background around every star, which is exactly what the subtraction is meant to preserve.

\subsubsection{Fitting and Subtraction Procedure}
\label{sec:methods-psf-fitting}

For each tile we select AllWISE sources with $W1 < 11$ falling within the footprint, convert their sky positions with the tile WCS, and reject any source whose position does not survive a round trip through the distortion solution to better than $1\arcsec$. The Level-2 astrometric solution is a third-order SIP polynomial supplied with its own inverse, and over a grid of $3.4\times10^{5}$ positions spanning the array on a sample of tiles the round-trip residual has a median of $10^{-10}$~arcsec; for the $1005$ catalogued sources with $W1<11$ on the tile of Fig.~\ref{fig:tile-overview} the largest residual is $3\times10^{-6}$~arcsec. The test is a sanity check against a corrupted solution rather than an active selection, and no source in this work has been removed by it. The magnitude limit is empirical: fitting fainter sources begins to remove genuine small-scale structure rather than stars, and what lies below the limit is handled by the median filter of Sect.~\ref{sec:methods-median} instead.

At each source position the tile is modelled as
\begin{equation}
  I \simeq a\,P + c_x\frac{\partial P}{\partial x} + c_y\frac{\partial P}{\partial y} + c_r\left(x\frac{\partial P}{\partial x} + y\frac{\partial P}{\partial y}\right) + b + g_x x + g_y y ,
\end{equation}
where $P$ is the PSF of Sect.~\ref{sec:methods-psf-construction}. The two first-derivative terms absorb sub-pixel errors in the catalogue position without refitting the centroid; the radial-moment term $c_r$ absorbs any residual isotropic mismatch in PSF size; and the plane $b + g_x x + g_y y$ represents the local diffuse background, which is fitted simultaneously so that the amplitude $a$ does not absorb the very emission we are trying to preserve. The fit is a weighted least-squares solution in which pixels inside an annulus around the star are upweighted by a factor of 50, so that the solution is driven by the wings, where the contamination of the surrounding diffuse background is set, rather than by the bright core. It is iterated with asymmetric sigma clipping ($3\sigma$ above the model, $6\sigma$ below), which rejects neighbouring sources without rejecting the negative excursions of the diffuse background itself.

Two regimes are distinguished by catalogue brightness. For bright stars ($W1 < 5$) the upweighted annulus runs from 12 to 65~pixels ($1.2\arcmin$--$6.7\arcmin$), the saturated core inside $16.5$~pixels is excluded from the fit, so that in practice the solution is set by the annulus from 16.5 to 65~pixels, the radial-moment term is included, and a radial stretch of the PSF in the range $0.79$--$1.26$ is optimised before the linear fit. The stretch is not a cosmetic degree of freedom: the model is built at the band centre, whereas any given star is observed at some other wavelength within Detector~4, and the stretch recovers the corresponding change in PSF width. For faint stars ($5 \le W1 < 11$) the annulus is narrow (3--20~pixels), the core is included, and neither the stretch nor the radial-moment term is used, since neither is constrained at that signal-to-noise. Faint stars lying within 10~pixels of a bright or ultrabright star are not fitted at all, because their local background plane would be set by their neighbour's wing rather than by the sky.

Two safeguards act against over-subtraction. First, the fitted amplitude is compared with the 20th percentile of the data-to-model ratio measured across the inner wings; if it exceeds that value by more than 30\%, the percentile value is adopted instead. Second, only positive amplitudes are subtracted. Sources whose fit returns a non-positive amplitude or fails to converge are left in place for the median filter. Stars brighter than $W1=5$ have their saturated core ($r<16.5$~pixels, $1.7\arcmin$) set to NaN after subtraction, since no model can recover information that the detector did not record.

\subsubsection{Treatment of Ultrabright Stars}
\label{sec:methods-ultrabright}

About two hundred stars are bright enough that their scattered-light wings extend far beyond the $80$-pixel footprint over which the standard model is defined, and they lie far beyond the brightness to which WISE photometry of saturated sources remains reliable, $W1\approx2$ \citep{Cutri2012}, so that a $W1$ ranking alone misses some of the objects that most need an extended model. We therefore define an ultrabright sample as the union of two lists: the brightest AllWISE sources by $W1$, and the brightest entries of the Yale Bright Star Catalogue \citep{Hoffleit1991}, the latter ranked not by apparent visual magnitude but by a blackbody estimate of their brightness at the band centre, obtained from the $B-V$ colour through the blackbody-based colour--temperature relation of \citet{Ballesteros2012}, which serves here only to order the sample and, like any estimate from $B-V$, is least reliable for the hottest stars, and a Planck ratio between $V$ and $3.12$~\micron. This matters in both directions: a cool giant such as Betelgeuse is modest in $V$ ($V=0.5$) but is the brightest infrared source in the sample, while a hot star bright in $V$ can be unremarkable at $3$~\micron.

Two different samples follow from this, and they should not be confused. The model is \emph{built} from the 250 brightest entries of each list, merged to 432 distinct stars after removing optical entries within $10\arcmin$ of an AllWISE one. The stars that are \emph{subtracted} from every tile are the 100 brightest of each list, merged in the same way to 197 stars, whose AllWISE members span $-2.00 < W1 < -1.81$ and whose optical members span an estimated $2.6 < m_{3.12} < 6.7$.

A second PSF is stacked from this sample on an $851\times851$~pixel ($\pm0.73\degr$) footprint, with the same recentring, wavelength rescaling and median combination as before. Because two ultrabright stars closer than $500$~pixels would each contaminate the other's stamp, both members of such a pair are dropped. The saturated region is correspondingly larger, so the power law is fitted between 14 and 60~pixels instead, returning $\alpha=2.78$ over that range; the Gaussian core and the pedestal subtraction are otherwise those of Sect.~\ref{sec:methods-psf-construction}.

These stars are subtracted first, before the standard bright and faint loop, using the same seven-parameter treatment with an upweighted annulus running from 30 to 300~pixels. The ordering is not incidental: any ordinary star sitting on an unsubtracted ultrabright wing would have its local background plane biased by that wing, so the ultrabright stars are matched out of the standard catalogue by sky position and removed before the rest of the tile is fitted.

The two stacks of Fig.~\ref{fig:psf-model} share the eight symmetric diffraction spikes, but the wider ultrabright footprint also reveals structure the standard stack is too small to contain. A compact blob appears near $x\approx+3\arcmin$, $y\approx25\arcmin$, with the appearance of a faint reflection within the optics rather than of anything on the sky. The standard stack in turn shows two faint rings, roughly 10~pixels across, centred near $y\approx\pm63$~pixels at either horizontal end of the frame. We have not identified either feature.

The aim for these objects is to remove the extended wings that no mask of reasonable size could cover, not to model the stars completely: their wings are shaped by the optics in ways that differ from star to star, and a model per object is left to future work. The core, where the model is least reliable, is therefore masked out to $r=200$~pixels ($20.5\arcmin$), both during the subtraction and in the reliability mask (Sect.~\ref{sec:postproc-mask}). What the subtraction buys is the sky outside that radius, which would otherwise be contaminated out to degree scales.

\subsubsection{Median-Filter Removal of Residual Faint Sources}
\label{sec:methods-median}

The explicit fit only removes catalogued sources above the magnitude limit. What is left, the sources below that limit and the small number whose fit was rejected, is removed statistically with a NaN-aware $11\times11$~pixel ($1.1\arcmin$) median filter applied to the star-subtracted tile. A median returns the local background wherever the contaminating source is smaller than the kernel, so point-like residuals are suppressed while extended emission is preserved. SPHEREx undersamples its point-spread function, whose diffraction core is narrower than a native pixel (Sect.~\ref{sec:methods-psf-construction}), so the residual of a star below the magnitude limit occupies only a few of the 121 pixels of the kernel. The kernel size sets the scale on which this stops being true, since structure comparable to the kernel is attenuated along with the stars. The filter is applied to the tile at its native resolution, before the smoothing of Sect.~\ref{sec:methods-smoothing}, so the relevant comparison is between the kernel and the resolution of the final maps rather than that of the data on which it acts. At $1.1\arcmin$ the kernel is an order of magnitude larger than the instrumental PSF and well below the $2.6\arcmin$ resolution of those maps, so the diffuse structure it attenuates is structure the final maps do not resolve in any case. Its contribution to the effective beam is accounted for in the smoothing step of Sect.~\ref{sec:methods-smoothing}.

\begin{figure*}[t]
  \centering
  \includegraphics[width=\textwidth]{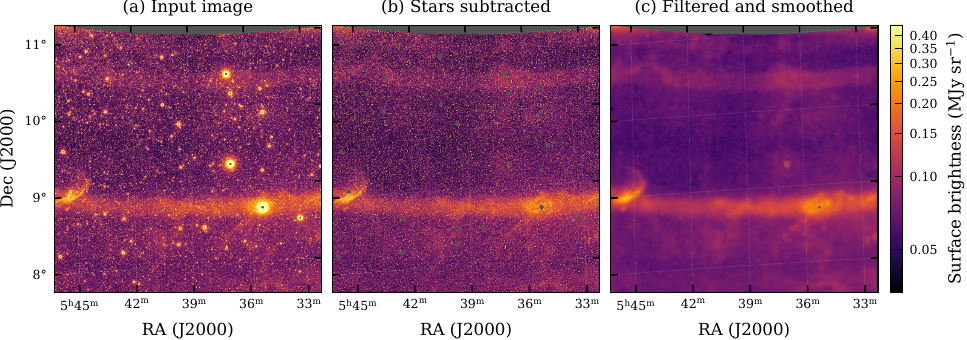}
  \caption{The per-tile processing chain, applied to the Detector-4 tile of Fig.~\ref{fig:tile-overview}. \textbf{(a)} After flag masking and zodiacal-light subtraction, the input to the stellar subtraction. \textbf{(b)} After fitting and subtracting the empirical PSF at every catalogued source above the magnitude limit, with the saturated cores of the brightest stars masked. \textbf{(c)} After the median filter and the Gaussian smoothing to $2.6\arcmin$, which is what is coadded into the HEALPix maps. All three panels share a single histogram-equalised colour scale.}
  \label{fig:star-subtraction}
\end{figure*}

Pixels that were masked earlier, whether by the flag mask or by the saturated-core mask of the stellar subtraction, are filled before the filter is applied, by propagating the median of their valid neighbours inward shell by shell. This prevents masked regions from growing at each subsequent step and from being spread further by the smoothing. The interpolated area is dominated by the saturated cores of the stars brighter than $W1=5$, whose positions are known. Rather than masking all of these cores, the residual-bright-star tier of the reliability mask (Sect.~\ref{sec:postproc-mask}) tests each one against its surroundings and flags it, at the same $16.5$-pixel ($1.7\arcmin$) radius, only where the filled core departs from them, so that a filled core indistinguishable from the surrounding sky is kept. Figure~\ref{fig:star-subtraction} shows the result of the stellar subtraction and the median filter on one tile, together with the smoothing of Sect.~\ref{sec:methods-smoothing}.

\subsection{Regridding and Full-Sky Accumulation}
\label{sec:methods-healpix}

The processed tiles are combined into a spectral data cube on the sky: one full-sky HEALPix map per narrowband channel, all on the same pixels, which the spectral extraction of Sect.~\ref{sec:methods-pah} then fits pixel by pixel. Each tile is first smoothed to a common angular resolution and then accumulated channel by channel.

\subsubsection{Smoothing to a Fixed Angular Resolution}
\label{sec:methods-smoothing}

The median filter of Sect.~\ref{sec:methods-median} is defined in detector pixels, whereas the maps must be delivered at a single, known angular resolution. We therefore convolve each tile with a Gaussian whose width in pixels is computed from that tile's own astrometric solution.

The target resolution follows from the pixelisation of the output maps. At $N_{\rm side}=4096$ the HEALPix pixel subtends $0.86\arcmin$ and the harmonic band limit of the map is $\ell_{\rm max}=3N_{\rm side}-1=12287$. A Gaussian beam is not strictly band-limited, so a resolution is lossless for this pixelisation only in the sense that its transfer function, $B_\ell=\exp[-\ell(\ell+1)\sigma^{2}/2]$, has become negligible by $\ell_{\rm max}$. Requiring a suppression of $10^{3}$ at $\ell_{\rm max}$ gives $\sigma_{\rm min}=\sqrt{2\ln 10^{3}}/\ell_{\rm max}$, that is ${\rm FWHM}_{\rm min}=2.45\arcmin$, or $2.9$ times the pixel size. This is the same condition as the rule of thumb that the pixel size should not exceed a third of the beam, used in CMB map-making for the same reason. We therefore adopt $2.6\arcmin$, close to the finest resolution this pixelisation supports.

The Gaussian kernel needed to reach that resolution cannot be obtained from the usual quadrature relation, because the median filter of Sect.~\ref{sec:methods-median} is nonlinear and has no closed-form transfer function. We calibrate it by forward modelling instead. A synthetic sky is built from a power-law diffuse background ($P(k)\propto k^{-2.5}$), filaments of $0.4$--$3.0\arcmin$ width, compact cores of $0.5$--$2.0\arcmin$ width and unsubtracted point sources at the level expected below the magnitude limit; it is passed through the real median filter followed by a Gaussian of trial width, and compared against the same sky convolved directly to $2.6\arcmin$ with no filtering. The trial width minimising the residual is $2.44\arcmin\pm0.02\arcmin$ over eight independent realisations, some $6\%$ below the naive value, and this is the kernel we apply.

The same simulations tell us how much of a real feature survives the filter. We measure this end to end, comparing each injected feature in the delivered map with the same feature smoothed directly to $2.6\arcmin$ without filtering, so that the slightly narrower $2.44\arcmin$ kernel, which partly makes up for what the median removes, is included. Features at least as wide as the map resolution come through intact: a compact feature of $2.6\arcmin$ keeps its peak to better than 1\% and 95--100\% of its flux, depending on how much it stands out from its surroundings, and one twice as wide keeps more than 99\%. Filaments, which the median sees as flat along their length, keep more than 99\% at this width. Only features narrower than the beam lose flux: an isolated feature of $1.1\arcmin$, the size of the kernel itself, keeps 73--96\%, and point-like structure is removed entirely, which is what the filter is for. The diffuse emission on the scales these maps resolve is therefore transmitted essentially unaltered.

\subsubsection{Full-Sky HEALPix Accumulation}
\label{sec:methods-healpix-accum}

Each processed tile is projected channel by channel. Every valid detector pixel has a sky position and a wavelength, which give a HEALPix index and a channel index; the pixel then contributes $I/\sigma^{2}$ to a flux accumulator and $1/\sigma^{2}$ to a weight accumulator for that cell. The science map and its variance are recovered at the end as $\sum(I/\sigma^{2})/\sum(1/\sigma^{2})$ and $1/\sum(1/\sigma^{2})$, which is the inverse-variance-weighted mean and its variance, where $\sigma^{2}$ is the per-pixel variance of the Level-2 VARIANCE plane \citep{Akeson2026}. A count of contributing exposures is kept per channel and released alongside the maps.

Tiles are found by stepping through the pixel centres of an $N_{\rm side}=256$ HEALPix grid ($13.7\arcmin$ spacing), and claiming at each grid point the nearest tile not yet used. Where coverage is dense, more tiles overlap a grid point than can be claimed this way; every tile returned by the lookups but never claimed is processed in a final pass, so that every tile covering the footprint is accumulated, and none more than once.

The practical limitation at QR2 depth is coverage rather than noise. Most lines of sight have been visited only a few times, and each visit contributes to a given narrowband channel only over the strip of the array where the filter passes that wavelength. Per-channel coverage is therefore not yet uniform, and small-scale gaps remain. Since the accumulator is additive and resumable by construction, the same maps can be refilled as more of the mission accumulates. One region is missing for another reason: as in \citet{Murgia2026}, the central longitudes of the Galactic plane have no usable Level-2 data in this release, because the astrometric solutions of fields that crowded are not yet reliable. This is expected to be resolved by the QR3 SPHEREx release rather than by further accumulation of the present one.

\subsection{PAH Spectral Extraction}
\label{sec:methods-pah}

\begin{figure}[t]
  \centering
  \includegraphics[width=\columnwidth]{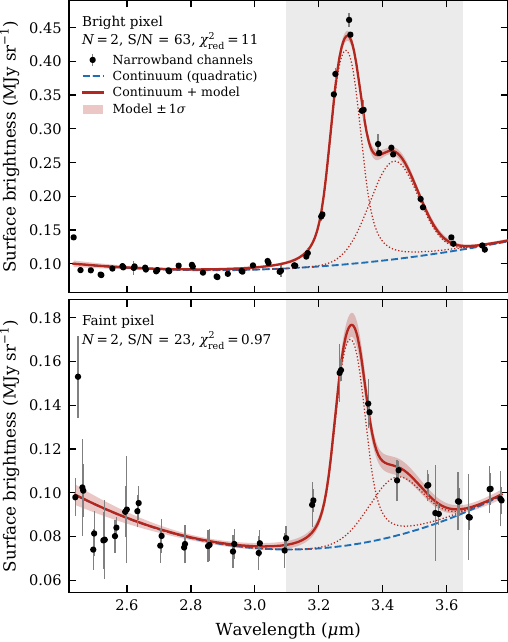}
  \caption{The per-pixel extraction at a bright pixel near Barnard~30 (top) and at a fainter one near Meissa ($\lambda$~Orionis) (bottom). Points are the narrowband channel values with their propagated uncertainties, plotted at their corrected wavelengths (Sect.~\ref{sec:methods-pah-wavelengths}), the dashed line is the fitted continuum, and the solid line is the winning continuum plus line model, with its individual components dotted; the 3.3~\micron\ component is the instrument-convolved Drude of Sect.~\ref{sec:methods-pah-fitting} and the 3.4--3.5~\micron\ one a Gaussian. The grey region marks the line window, which is withheld from the continuum fit and is the only range in which a component may be centred, and the red shading is the $1\sigma$ uncertainty of the fitted model itself, propagated from the covariance of its parameters. At the peak of a bright band the model sits slightly under the data. This is expected: the adopted Drude at the intrinsic width taken from the observed band of \citet{TokunagaBernstein2021} is marginally broader than the band itself (Sect.~\ref{sec:methods-pah-resolution}), so the amplitude that minimises $\chi^{2}$ across the whole profile falls a little short at the single brightest channel while matching the flanks. The larger reduced $\chi^{2}$ of the bright pixel reflects the channel-to-channel calibration floor of Sect.~\ref{sec:results-map}, which dominates for bright pixels such as this one.}
  \label{fig:pah-spectra}
\end{figure}

The coadded Detector-4 channel maps (Sect.~\ref{sec:methods-healpix}) give every sky pixel a sampled spectrum spanning $2.42$--$3.82$~\micron, from which we extract the integrated radiance of the 3.3~\micron\ aromatic C--H stretch and of the neighbouring 3.4--3.5~\micron\ aliphatic complex. Because the wavelength solution is per detector pixel rather than per tile (Sect.~\ref{sec:data-tiles}), stacking the channel maps at fixed sky position yields a genuine, if coarsely sampled, spectrum. This is the essential difference between the map presented here and broadband PAH proxies such as WISE~12~\micron: the band emission is separated from the underlying continuum at the level of the spectrum itself, so stellar and continuum emission enter the fit as a baseline to be removed rather than as an irreducible contaminant of the photometry. The fit described below is performed independently at every pixel, and Fig.~\ref{fig:pah-spectra} shows it at two representative ones.

\subsubsection{Continuum and Line Model}
\label{sec:methods-pah-fitting}

Channels within a line window of $3.10$--$3.65$~\micron\ are withheld from the continuum fit and are the only wavelengths at which a line component may be centred. The red edge is set well past the 3.3~\micron\ feature because the aliphatic complex does not return to the continuum until $\sim$3.6~\micron; anchoring a baseline at 3.6~\micron, as two-point band-excess methods do \citep[e.g.][]{Murgia2026}, therefore places a continuum reference inside genuine hydrocarbon emission and biases the recovered band flux low. Continuum channels blueward of $2.60$~\micron\ are also excluded.

The surviving line-free channels are fit with three variance-weighted candidate continua: a straight line, a second-order polynomial and a power law $I=A\lambda^{\alpha}$, and the model minimising the Akaike Information Criterion \citep[AIC;][]{Akaike1974} is subtracted. Over a wavelength range spanning only a factor $\sim$1.6 these forms are close to degenerate, and we make no claim that the selected one is physically preferred; the AIC comparison simply prevents a single imposed shape from absorbing real curvature in the baseline. In practice that curvature is usually present: the quadratic is selected at $76$\% of the pixels, and at $94$\% of those where the band is detected at $S/N\geq10$, the straight line at $16$\% and the power law at $8$\%, both mostly in faint sky where the three forms cannot be told apart.

The continuum-subtracted residual is then modelled as a sum of up to two line components, fit by variance-weighted least squares with \texttt{lmfit} \citep{Newville2014}. The uncertainties of the fitted parameters, and therefore those of the released radiances, which are propagated from them, are the covariance of this fit multiplied by its reduced $\chi^{2}$. We keep this scaling deliberately: the spectral model, particularly for the baseline, is likely less complex than the data, and the radiance uncertainties should reflect how well it actually describes each spectrum rather than the nominal channel variances alone. The absolute scale of the channel variances therefore cancels, only their relative weights matter, and the uncertainties follow the actual scatter of the channels about the model, whatever its origin: noise that the propagated variances misjudge, calibration and zodiacal-light offsets between the exposures that fill different channels, or an imperfect spectral model. We fit $N=1$ (the 3.3~\micron\ feature alone) or $N=2$ (the 3.3~\micron\ feature plus the aliphatic complex) at every pixel and adopt whichever has the lower AIC, so the data decide per pixel whether a second component is justified. Where the band is detected, $N=2$ is selected in 80\% of pixels. A pixel is fit only if it has at least six valid channels, of which at least four lie in the continuum region and three in the line window.

Two properties of these uncertainties matter for their use. First, their scale is set empirically, by the $\chi^{2}$ scaling, rather than propagated: the channel variances are those of the native Level-2 pixels, accumulated as in Sect.~\ref{sec:methods-healpix-accum} but not propagated through the median filter and the smoothing, so their absolute value does not describe the smoothed maps and only their relative weights enter. Second, they are per pixel. The maps are sampled at $0.86\arcmin$ with a $2.6\arcmin$ beam, whose solid angle covers about ten pixels, so neighbouring pixels are correlated, and the uncertainty of a regional average falls with the number of beams it contains rather than with the number of pixels. The scaling is examined against the brightness of the sky in Sect.~\ref{sec:results-map}; a direct test against the scatter between maps built from disjoint subsets of the exposures, which the additive accumulator of Sect.~\ref{sec:methods-healpix-accum} makes possible, is left to the next release.

The 3.3~\micron\ band is fitted with a Drude profile convolved with the instrumental response of Sect.~\ref{sec:methods-pah-resolution}, not with a Gaussian. PAH band profiles are Drude-like on physical grounds, and because the band is only marginally resolved the profile has to be imposed rather than recovered: a pseudo-Voigt fit to the stacked footprint spectrum returns a Lorentzian fraction consistent with either limit, and the two forms differ by $\Delta\mathrm{AIC}=1.9$, so goodness of fit cannot choose between them. Fixing the intrinsic width at the measured value of \citet{TokunagaBernstein2021}, as Sect.~\ref{sec:methods-pah-resolution} sets out, also leaves the band with one free parameter fewer than a Gaussian would have, only its amplitude and centre being fitted.

The choice is not free of consequences, and they fall on the split rather than on the sum. A Drude has wings a Gaussian does not, and its red wing falls exactly where the aliphatic complex sits, so fitting the wrong one of the two does not show as a poor fit: it moves flux between the components. Refitting with an all-Gaussian model moves the released radiances by $+6.0$\% in the 3.3~\micron\ component and $-10.7$\% in the 3.4--3.5~\micron\ component, but by only $+1.7$\% in the total. We carry this as a profile-shape systematic.

One contaminant cannot be removed by any choice of window. The hydrogen Pfund-$\delta$ recombination line ($9\rightarrow5$) lies at $3.2970$~\micron\ in vacuum, within $0.01$~\micron\ of the aromatic band centre and far inside the instrumental width, so in ionised gas it is blended with the feature and counted as band radiance. Its strength is nevertheless tied to that of Pfund-$\gamma$ ($8\rightarrow5$) at $3.7406$~\micron\ almost independently of conditions, because the ratio of two recombination lines of the same series is set by atomic physics and depends only weakly on the temperature and density of the gas. In the Case~B limit \citep{BakerMenzel1938}, with the line emissivities of \citet{Storey1995}, $I(\mathrm{Pf}\text{-}\delta)/I(\mathrm{Pf}\text{-}\gamma)=0.692$ at $T_{\rm e}=10^{4}$~K and $n_{\rm e}=10^{2}$~cm$^{-3}$ and varies only between $0.680$ and $0.702$ over $T_{\rm e}=5000$--$20\,000$~K and $n_{\rm e}=10^{2}$--$10^{4}$~cm$^{-3}$ \citep[computed with PyNeb;][]{Luridiana2015}, and Case~A gives $0.690$ under the same conditions, which is what makes the blend boundable at all.

Pfund-$\gamma$ is not in clean sky either. It falls inside the $3.65$--$3.83$~\micron\ interval that supplies the only continuum channels redward of the line window, so the baseline is fitted straight through it and part of the line is absorbed before it is measured. What we recover is therefore a lower bound on Pfund-$\gamma$, and the contamination inferred from it a lower bound on the contamination. With that caveat the effect is small. Injection tests recover a known line without bias over two orders of magnitude in strength and return zero when none is present, and stacking 4000 pixels drawn from across the footprint, outside the reliability mask of Sect.~\ref{sec:postproc-mask}, detects Pfund-$\gamma$ at $93\sigma$, implying a mean Pfund-$\delta$ contamination of the integrated band radiance of at least $1.31\pm0.01$\%. It is not uniform, and it is weakest where the band is brightest: the same stack restricted to the 400 brightest pixels gives $0.25$\%, because the band radiance rises more steeply between faint and bright sightlines than the recombination emission does. The $1.3$\% figure is an average over the released sky rather than a property of any one sightline, and we treat it as a known additive bias rather than correcting for it pixel by pixel.

Both problems have one cause, which is that Detector~4 ends at $3.82$~\micron, before the emission does, leaving no line-free continuum redward of either the aliphatic complex or Pfund-$\gamma$. Extending the spectra onto Detector~5 ($3.82$--$4.42$~\micron) would anchor the continuum redward of the complex and place Brackett-$\alpha$ in the same fit, so that the recombination series could be subtracted rather than bounded. We leave that to future work.

\subsubsection{Spectral Resolution and Component Priors}
\label{sec:methods-pah-resolution}

The spectral resolution of the maps is set by the LVF and not by our channel binning: the 102 logarithmically spaced channels of Sect.~\ref{sec:data-band4} correspond to $R\approx220$ and oversample an instrumental response whose design value is $R=35$ \citep{Crill2020,Hui2026}. Both features are therefore at best marginally resolved, and the fit has to be told what they look like rather than asked to discover it. The order matters: the aromatic band is held rigid, so any error in the assumed response has only one place to go, namely the width of the second component. We therefore fix the response first, from an external measurement, and measure the bands against it afterwards.

SPHEREx characterised the transmission of every detector pixel in the laboratory and releases the resulting per-channel spectral response functions \citep{Hui2026}. We take the Detector-4 curves and add to them the further broadening our own processing introduces: the filter encodes wavelength as position along the dispersion axis, so the $2.6\arcmin$ smoothing of Sect.~\ref{sec:methods-smoothing} projects onto wavelength as a further $0.0215$~\micron. At $3.295$~\micron\ the resulting response has a FWHM of $0.0937$~\micron, implying $R=35.2$, with a median of $37.4$ over the seventeen characterised channels spanning Detector~4, and its shape is Gaussian rather than rectangular. Refitting 7592 footprint pixels under a grid of candidate responses returns the same value, so the sky and the laboratory agree.

With the response fixed, the aromatic band can be measured against it, and its centre is the well-determined quantity. Fitted over 4000 stratified footprint pixels with the priors switched off, the band returns a centroid of $3.2845$~\micron, $0.0042$~\micron\ from the mean band centre of $3.2887\pm0.0009$~\micron\ measured on nine Galactic sources by \citet{TokunagaBernstein2021} and so under a twentieth of the instrumental FWHM from it. We adopt that value as a soft Gaussian prior with $\sigma_{\rm prior}=0.008$~\micron, about a fifth of the instrumental width: wide enough that the fit can follow any real shift, narrow enough that a noise-dominated pixel cannot move the component off the feature. Hard bounds of $3.27$--$3.31$~\micron\ stop it leaving the band altogether.

Its width is not fitted: the Drude carries the FWHM of $0.041$~\micron\ measured by \citet{TokunagaBernstein2021} and the response carries the rest. That width is itself an observed one, so adopting it as intrinsic is mildly conservative. These data cannot constrain it: the response accounts for about two-thirds of the observed variance, and subtracting it magnifies small errors: $\sigma=0.0437$~\micron\ observed against $0.0361$~\micron\ from the response leaves an intrinsic $\sigma$ of $0.025$~\micron, and a 5\% error in the first becomes a 16\% error in the last. Because the width is imposed, the observed width becomes a prediction. Convolving the measured response with a band of the adopted width and refitting the result exactly as the data are fitted gives $\sigma=0.0392$~\micron\ for a Gaussian intrinsic profile and $0.0475$~\micron\ for a Drude, against $0.0437$~\micron\ observed. The observation falls between the two, as it should if the true profile does, which is what the pseudo-Voigt of Sect.~\ref{sec:methods-pah-fitting} also indicates. The adopted Drude is the wider of the two, and a model band slightly too wide sits below the brightest channel while matching the flanks, which is what the top panel of Fig.~\ref{fig:pah-spectra} shows. Letting the intrinsic width float would close that gap only by absorbing the profile into the width: the same observed $0.0437$~\micron\ implies an intrinsic FWHM of $0.029$~\micron\ for a Drude and $0.063$~\micron\ for a Gaussian, a factor of two apart and bracketing the adopted $0.041$~\micron. A fitted width would therefore restate the assumed profile rather than measure the band, the circularity that measuring the response was meant to end.

The 3.4--3.5~\micron\ feature is not a single transition but a blend of methyl and methylene C--H stretching modes, and its priors follow from that structure rather than from a fit. The sub-peaks are documented in diffuse-ISM absorption at $2955$, $2925$ and $2870$~cm$^{-1}$ ($3.384$, $3.419$ and $3.484$~\micron) \citep{Sandford1991,Pendleton1994,PendletonAllamandola2002}; in emission from photodissociation regions the complex also carries weaker features near $3.51$ and $3.56$~\micron\ \citep{Bernstein1996}, and \citet{Boersma2026} separate a single 3.4~\micron\ component from the 3.3~\micron\ band in SPHEREx data of a bright region of this kind. Placing narrow lines at those positions, convolving them with the measured response and passing the result through the same fitting code used on the data predicts what this pipeline should report for a complex of a given composition. Table~\ref{tab:aliphatic-forward} collects those predictions, and both the centre and the width are governed by the same quantity: how much of the flux sits redward of $3.47$~\micron.

\begin{table}
\caption{Expected centres and widths for a forward-modelled 3.4--3.5~\micron\ complex. Narrow lines at the documented sub-peak positions are convolved with the measured instrumental response and refitted by the pipeline. The rows differ only in how strongly the two red emission features at $3.51$ and $3.56$~\micron\ are weighted against the three absorption sub-peaks; $f_{>3.47}$ is the resulting fraction of the input radiance lying redward of $3.47$~\micron, and $\lambda_{0}$ and $\sigma$ are the centre and width the pipeline recovers. The last row is the measurement on the sky, made with the priors switched off.}
\label{tab:aliphatic-forward}
\centering
\begin{tabular}{lccc}
\toprule
Weighting & $f_{>3.47}$ & $\lambda_{0}$ & $\sigma$ \\
 & & (\micron) & (\micron) \\
\midrule
Absorption sub-peaks only  & 0\%  & 3.421 & 0.046 \\
With the red features      & 39\% & 3.428 & 0.069 \\
Red features doubled       & 49\% & 3.445 & 0.070 \\
All five equally weighted  & 60\% & 3.476 & 0.079 \\
\midrule
Measured                   & --   & 3.440 & 0.081 \\
\bottomrule
\end{tabular}
\end{table}

The centre is measured on the 1553 pixels, of 12\,000 drawn across the footprint, that carry at least 92 of the 102 channels and whose band signal-to-noise exceeds 20, and it returns $3.4402$~\micron. Theory meets the measurement there: a complex with roughly half its flux redward of $3.47$~\micron\ predicts $3.445$~\micron, within $0.005$~\micron, and the weighting that does it is the one including the red emission features at the strength photodissociation regions show. We adopt a centre of $3.44$~\micron\ with $\sigma_{\rm prior}=0.01$~\micron, bounded by the span of the documented sub-peaks. That $\sigma_{\rm prior}$ is neither the spread of the predictions in Table~\ref{tab:aliphatic-forward} nor the pixel-to-pixel scatter, most of which is measurement error, but how much the centre genuinely varies across the sky: the scatter of the per-region medians over 13 well-sampled $N_{\rm side}=2$ superpixels, with each median's own sampling error removed in quadrature, is $0.0086$~\micron.

The component's width follows on the same pixels and agrees less closely, but still without having been tuned. The measurement gives $\sigma=0.0812$~\micron\ against the $0.070$~\micron\ predicted by the weighting that matched the centre, a 16\% difference. That difference is not an artefact of holding the aromatic band rigid: stepping its assumed intrinsic width across the plausible range moves the fitted aliphatic width the other way, at $\mathrm{d}\sigma/\mathrm{d(FWHM)}=-0.8$, so an aromatic band closer to the one the data prefer would widen the gap rather than close it. We adopt $\sigma=0.081$~\micron\ with $\sigma_{\rm prior}=0.008$~\micron. Between regions the component's width varies by $0.0043$~\micron, half the figure for its centre, and we set $\sigma_{\rm prior}$ proportionally further above that than we did for the centre because this is the model's only free width and therefore where any residual error in the instrumental response will collect. The upper bound of $0.13$~\micron\ sits well above anything measured, so it regularises the fit without ever constraining it.

This second component is named for its wavelength rather than for its carrier, because it integrates everything lying between the aromatic band and the red edge of the line window. In emission that interval carries anharmonic and hot-band emission from the 3.3~\micron\ mode itself, shifted to longer wavelengths and not described by a single Drude at the band centre \citep{Barker1987,Joblin1996}, alongside the aliphatic C--H stretches, and Detector~4 alone cannot separate them.

The three released products should therefore be trusted differently. Refitting the footprint under each modelling choice the data leave open --- a looser aliphatic prior, both components free, and the all-Gaussian profile --- moves the maps by up to 2\% in the combined radiance, 6\% in the 3.3~\micron\ component and 11\% in the 3.4--3.5~\micron\ component, the largest changes coming from the profile shape. These are ranges spanned under those choices rather than random errors, and we quote them alongside the statistical uncertainties rather than folding them in. Flux moved between the components leaves the combined radiance almost unchanged, so it is a measurement and the primary product of this work; the 3.3~\micron\ map is secure at the 10\% level; and the 3.4--3.5~\micron\ map is a model-dependent estimate rather than a measurement of aliphatic emission. We recommend the combined radiance wherever the science does not require the split.

\subsubsection{Wavelength Assignment}
\label{sec:methods-pah-wavelengths}

Each channel map collects the Level-2 pixels whose wavelength falls within that channel. The LVF maps wavelength onto position along curved lines of constant wavelength (Fig.~\ref{fig:tile-overview}), so on every tile a channel is a curved band about $2\arcmin$ wide, and each tile samples a given sky pixel at a single wavelength, anywhere within $\pm7$~nm of the channel centre. Fitting every sample at its channel centre therefore introduces a wavelength error that follows the tile geometry, repeating every $\sim2\arcmin$ across the lines of constant wavelength and jumping at tile edges. On a band as narrow and steep as the 3.3~\micron\ feature this error changes the fitted amplitude, and it imprints on the maps lines that follow the LVF curves, scale with the signal and are strongest in the split between the two components.

We therefore fit every sample at its own wavelength. The wavelength of a detector pixel follows from the lookup table of the spectral WCS, which is the same for all tiles, and each tile's celestial WCS places it on the sky, so the correction can be built without any tile data: averaging the offset from the channel centre over the $2.27\times10^{5}$ tiles that contribute to each sky pixel gives one full-sky wavelength-correction map per channel. The corrections move samples within their channels without shifting the band. With them the stacked spectra collapse onto a single band profile (Fig.~\ref{fig:pah-spectra} is drawn at the corrected wavelengths; at its bright pixel $\chi^2_{\rm red}$ falls from 16 to 11), the reduced $\chi^2$ falls in 75--83\% of pixels with signal-to-noise 15--80, and the mean radiances change by at most 1.6\%. This is a first-order correction: each pixel and channel receives the mean correction of the contributing tiles, with uniform weights rather than those of the accumulation, which is adequate at the present sparse coverage. A spectral treatment such as the interpolation of \citet{Cukierman2026}, which builds narrowband mosaics and interpolates along each mosaic pixel to a common target wavelength, would treat the spread of wavelengths within a channel explicitly, and we leave it to a future version. Mismodelling the band profile amplifies the same effect, because the lines arise from the mismatch between the band as modelled and as sampled: with the instrumental response modelled as a Gaussian instead of the measured response of Sect.~\ref{sec:methods-pah-resolution}, the lines in the combined radiance are several times stronger.

\subsubsection{Faint and Non-Detected Pixels}
\label{sec:methods-pah-lowsnr}

No significance threshold is applied anywhere in the extraction. Every pixel with enough valid channels is fit, and the resulting radiance is reported whatever its signal-to-noise; a pixel is left blank only when it has too few channels, never because its flux is faint or consistent with zero. This is deliberate. Over most of the sky the diffuse PAH signal is weak, and any acceptance cut would retain the positive half of a noise-dominated flux distribution while discarding the negative half, biasing the map high precisely where an unbiased estimate matters most for downstream stacking or cross-correlation.

For the same reason, component amplitudes are bounded below not at zero but at $-8\,\hat{\sigma}_{\rm amp}$, where $\hat{\sigma}_{\rm amp}$ is a per-pixel estimate of the amplitude noise obtained by linearised error propagation from the local channel variances at fixed centre and width. A hard positivity bound would force a true non-detection to pile up at, but never below, zero, the familiar non-negative least-squares bias; the adopted floor is loose enough never to clip a genuine noise fluctuation, and scales with the local noise level, while still keeping the optimiser bounded. A noise-only pixel is thus free to scatter symmetrically about zero, as a null result should.

\subsection{Destriping}
\label{sec:Post-processing}
\label{sec:postproc-destripe}

\begin{figure*}[t]
  \centering
  \includegraphics[width=\textwidth]{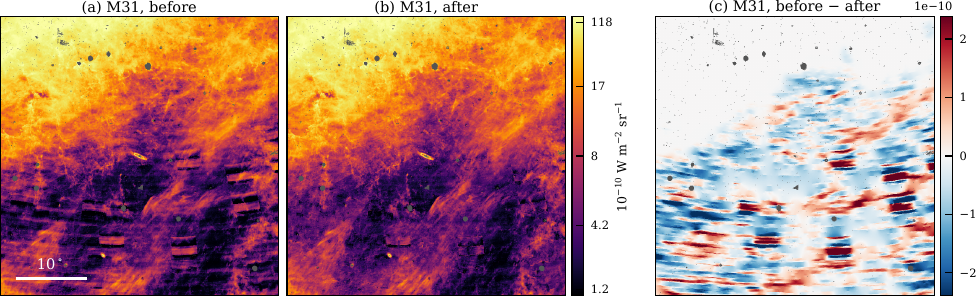}
  \caption{The combined radiance in a $40\degr$ field around M31, in a gnomonic projection aligned with the ecliptic so that the stripes run horizontally. \textbf{(a)} Before and \textbf{(b)} after destriping, both histogram-equalised on the mapping of (a); \textbf{(c)} their difference, which is the subtracted correction, on a diverging scale centred on zero. Pixels flagged by the recommended reliability mask (Sect.~\ref{sec:postproc-mask}) are grey.}
  \label{fig:destripe-m31}
\end{figure*}

Faint regions of the coadded maps are crossed by stripes of degree scale that run along lines of constant ecliptic latitude, the ecliptic parallels (Fig.~\ref{fig:destripe-m31}). They are an additive residual of the zodiacal light and of calibration offsets between overlapping tiles, which the per-pixel continuum absorbs only in part. Their size is set by the size of the Level-2 tiles. In ecliptic latitude, that is, perpendicular to their length, 92\% of their power lies on scales between $0.8\degr$ and $4.8\degr$, with maxima at $1.2$--$1.9\degr$ and $3.1$--$4.8\degr$, the first harmonic and the fundamental of the $3.5\degr$ tile. In ecliptic longitude, along their length, they are much longer: they run within a median $2.5\degr$ of the ecliptic parallels and stay coherent over many degrees, latitude profiles $0.5\degr$ apart in longitude correlating at 0.78, and still near 0.2 at $10\degr$, whereas sky of the same brightness decorrelates within about $2\degr$. They are spectrally smooth across the PAH line window and positively correlated between the two PAH components ($+0.42$), and the tile offsets that produce them are visible directly in the channel maps: neighbouring pixels covered by different tiles differ two to four times more than pixels covered by the same one, which corresponds to relative offsets of 3--4\% between tiles in each channel, as expected from zodiacal-light or calibration residuals. In the faint half of the footprint, where they can be measured directly, their amplitude does not depend on the sky level and is comparable to the sky signal itself.

Such stripes are best removed before the tiles are combined, by fitting the zodiacal light, or a per-tile offset and gradient, so that each tile matches its overlapping neighbours. For WISE 12~\micron\ \citet{MeisnerFinkbeiner2014} fitted a gradient to each exposure against the SFD 100~\micron\ map and then added the exposures one at a time, each with the single offset that best matched the stack built so far. That requires reprocessing the full Level-2 archive, about 16~TB, with the individual tiles retained through the spectral extraction. It is beyond the scope of this release and is planned for a future version of the maps; here we correct the coadded products, and release the unfiltered maps alongside the filtered ones.

The filter leverages the direction of the stripes: structure along the ecliptic parallels contains stripes and sky, whereas structure perpendicular to them contains sky alone and provides a null measured from the same data. It detects the edges of the stripes and rebuilds the stripes from them. On a $4\arcmin$ grid in ecliptic coordinates we compare each row of constant latitude with the next. A stripe edge appears as a step change in brightness that persists along the parallel, whereas the steps produced by sky do not, so we take a running median of the row-to-row differences along each parallel over $1.7\degr$ and keep only the steps that exceed twice the scatter of the same quantity measured perpendicular to the parallels. The correction is then the map whose row-to-row steps match the kept ones while staying as uniform as possible along each parallel, found in a single least-squares fit over the whole grid (a Poisson equation); fitting all rows together, rather than adding up the steps along each meridian, keeps the noise of individual steps from accumulating. Structure on scales above $10\degr$, much larger than any stripe, is removed from the correction, since there it can only come from small errors in the steps adding up over large distances. Where the sky is too structured for its steps to be trusted, the correction is instead interpolated along the parallels, and these regions are refined iteratively. The correction is subtracted from the full-resolution map, which is never resampled. It is computed independently for the combined and the aromatic radiance, and the filtered 3.4--3.5~\micron\ map is their difference, so the filtered products sum exactly. Figure~\ref{fig:destripe-m31} shows the result in a $40\degr$ field around M31.

The stripes matter only in faint sky, and that is where the filter is applied. In bright sky they amount to a few per cent of the signal, and cirrus whose latitude structure persists over $1.7\degr$ along a parallel is indistinguishable from a stripe edge, so filtering there would remove more sky than stripes. The correction is therefore tapered with sky brightness: its weight falls linearly from one to zero between the 50th and 75th percentiles of the $1\degr$-smoothed sky, and the weight map is smoothed over $1\degr$ so that the taper has no sharp edge, which lets the correction reach slightly beyond the 75th percentile. Above the 90th percentile, which includes the Galactic plane, the filtered and unfiltered maps agree to better than 0.1\%. In faint sky, the same limitation affects elongated structure lying close to an ecliptic parallel.

We calibrate the filter by forward modelling, as for the median filter of Sect.~\ref{sec:methods-smoothing}. We inject into the real map synthetic stripes with the measured amplitude ($9.5\times10^{-11}$~W~m$^{-2}$~sr$^{-1}$ rms), orientation and tile geometry, an isotropic Gaussian sky with $P(k)\propto k^{-2.5}$ at two to three times the structure of the real faint sky, and $6\degr$-long filaments, and recover each component by differencing runs with and without it. The real faint sky is somewhat flatter, with $P(k)\propto k^{-2.0}$ (16th--84th percentiles of the index $-2.4$ to $-1.6$) on these scales, but repeating the test with indices of $-2$ and $-3$ changes the fraction of faint sky removed at any scale by at most 0.5 percentage points. Applied without the taper, the filter removed up to 13\% of the injected bright sky while recovering only 18\% of the injected stripes there. With the taper it removes 88\% of the stripe variance in the faint half of the footprint and 87\% between the 50th and 75th percentiles, and at most 2.2\% of the injected sky at any scale from $0.5\degr$ to $10\degr$; the loss falls on structure within $30\degr$ of the ecliptic parallels (4--6\%), while structure perpendicular to them loses under 0.5\%. Between the 75th and 90th percentiles, which only the smoothed edge of the taper reaches, half of the stripes and 6\% of the sky are removed, and above the 90th neither. Statistics of the diffuse emission, such as power spectra or correlations with other tracers, are therefore preserved to a few per cent. Individual faint filaments aligned with the ecliptic parallels, however, lose up to a third of their flux, and should be measured on the unfiltered maps.

\section{Results and Discussion}
\label{sec:results}

\subsection{Released Products}
\label{sec:methods-pah-outputs}
For each pixel we report three integrated radiances with their propagated $1\sigma$ uncertainties: the combined radiance of the winning model, which is the primary map presented in this work, and the same quantity split into its 3.3~\micron\ aromatic component and the 3.4--3.5~\micron\ component redward of it, the latter populated only where $N=2$ was selected. The second is named by its wavelength rather than by its carrier, and is an estimate of the aliphatic emission, for the reason given in Sect.~\ref{sec:methods-pah-resolution}. The three are not equally robust: flux moved between the two components by a change in the model leaves the combined radiance almost unchanged, so the total is a measurement while the split is a model-dependent estimate, and the redward component, the smaller of the two, is the least secure of the three. Users who need only the band emission should take the total. Finally, we release per-pixel diagnostics: the selected $N$, the reduced $\chi^{2}$ and a channel coverage count, so that users can impose their own quality cuts downstream. The reliability mask of Sect.~\ref{sec:postproc-mask} is released as a single HEALPix file carrying one bit per condition, so that the conditions can be combined, loosened or ignored independently rather than only as the single recommended mask. The channel maps give the surface brightness per unit frequency, $I_\nu$, and the lines are fitted against wavelength, so each fitted component has an area $\int I_\nu\,{\rm d}\lambda$ in MJy\,sr$^{-1}$\,\micron. We convert it to the band radiance $\int I_\nu\,{\rm d}\nu=\int I_\lambda\,{\rm d}\lambda$ in W\,m$^{-2}$\,sr$^{-1}$ through ${\rm d}\nu=(c/\lambda^{2})\,{\rm d}\lambda$, with $c/\lambda^{2}$ evaluated at the centre of each component (3.29 and 3.44~\micron), which is accurate to about 1\% across these narrow bands, and release these band radiances, with the combined radiance as the sum of the two. Only the conversion depends on wavelength: the result is an energy per unit area, time and solid angle, directly comparable between bands and instruments, whereas the value of $\int I_\nu\,{\rm d}\lambda$ for a given energy depends on where the band lies. All radiances quoted in this paper are in these units. All products are rotated from the equatorial frame in which they are built to Galactic coordinates for release, and the destriped combined, aromatic and 3.4--3.5~\micron\ maps (Sect.~\ref{sec:postproc-destripe}) are released alongside the unfiltered ones. The maps, the reliability mask and an explanatory supplement describing the files and their known issues are available at \url{https://research.iac.es/proyecto/radioforegroundsplus/pages/data-products/spherex-pahs.php}.

\subsection{The Full-Sky PAH Maps}
\label{sec:results-map}

\begin{figure*}[!tp]
  \centering
  \includegraphics[width=\textwidth,height=0.84\textheight,keepaspectratio]{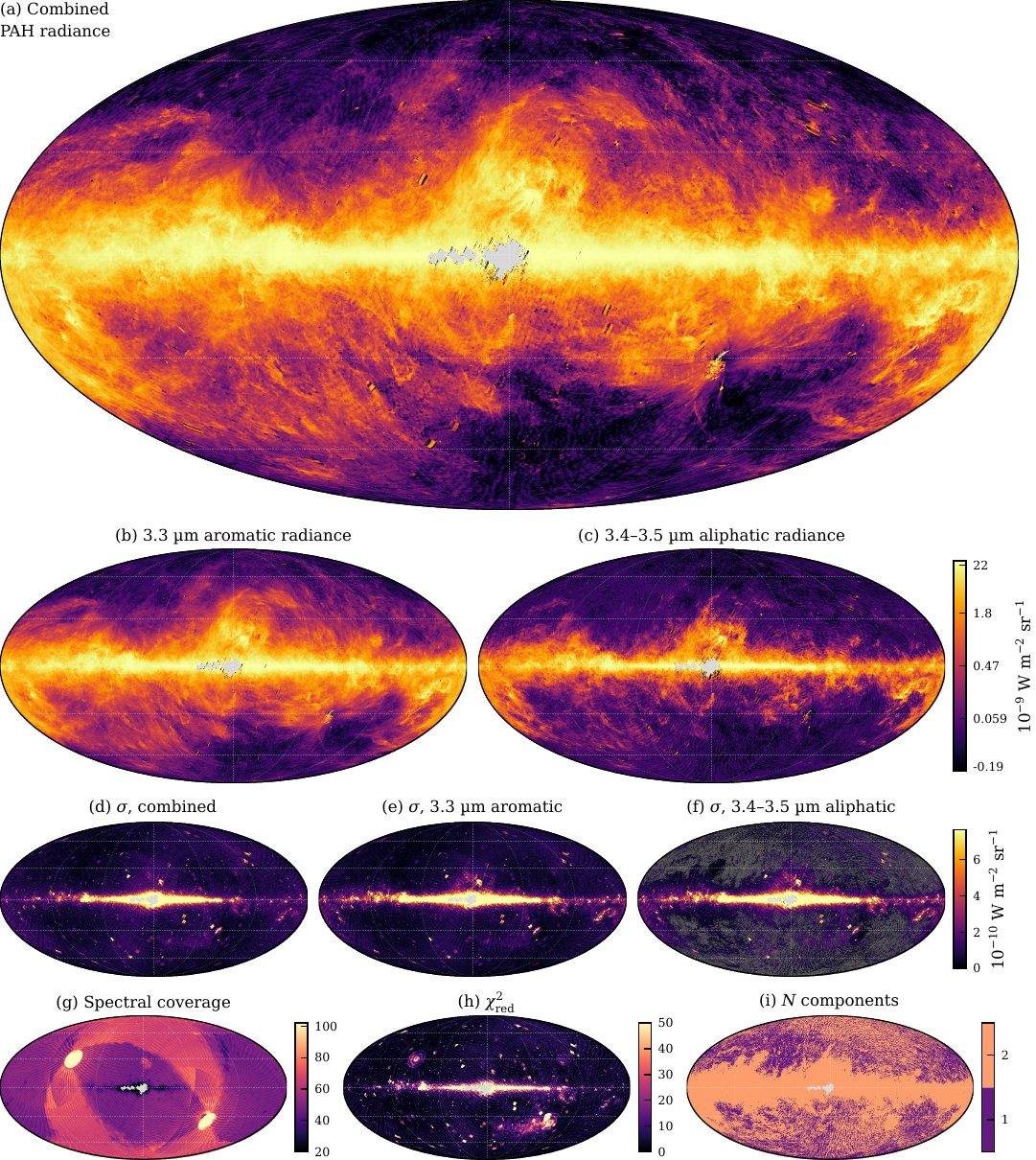}
  \caption{The released maps in Galactic coordinates. \textbf{(a)} Combined radiance of the 3.3 and 3.4--3.5~\micron\ bands. \textbf{(b)} 3.3~\micron\ aromatic and \textbf{(c)} 3.4--3.5~\micron\ aliphatic components; (c) is set to zero where only one component was fitted. Panels (a)--(c) show the destriped maps (Sect.~\ref{sec:postproc-destripe}) on a single histogram-equalised colour scale, taken from (a) and labelled in radiance to the right of (b) and (c). \textbf{(d)}--\textbf{(f)} The $1\sigma$ statistical uncertainties on (a)--(c), on the linear scale shown to the right of (f); destriping does not change them. \textbf{(g)} Number of valid channels, of 102. \textbf{(h)} Reduced $\chi^{2}$. \textbf{(i)} Number of line components fitted in each pixel. Light grey is unobserved sky. No reliability mask is applied.}
  \label{fig:fullsky-maps}
\end{figure*}

Figure~\ref{fig:fullsky-maps} presents the released maps, with the radiances shown after destriping (Sect.~\ref{sec:postproc-destripe}): the combined band radiance, the primary product of this work, in panel~(a); its 3.3~\micron\ aromatic and 3.4--3.5~\micron\ components in (b) and (c); their statistical uncertainties in (d)--(f); and the number of valid channels, the reduced $\chi^{2}$ and the number of line components fitted in (g)--(i). The present release covers $99.4$\% of the sky, or $41\,000$~deg$^{2}$, with at least 20 of the 102 channels; nearly all of the remaining 0.6\% lies within $5\degr$ of the Galactic plane. No reliability mask is applied in the figure, so that what the mask of Sect.~\ref{sec:postproc-mask} removes can be judged against the maps themselves.

The combined radiance is dominated by the Galactic plane, which the histogram-equalised colour scale compresses into its top end so that the fainter emission away from it remains visible. Off the plane the map recovers the extended emission of the local interstellar medium, from the high-latitude cirrus to the nearby molecular clouds and the photodissociation regions at their edges, as well as the brightest nearby galaxies, among them the Magellanic Clouds and M31. How closely this structure agrees with the broadband mid-infrared view is quantified in Sect.~\ref{sec:discussion-wise}. Along the faintest sightlines the band emission falls to the noise and the map scatters about zero, with negative pixels retained by construction (Sect.~\ref{sec:methods-pah-lowsnr}); this is also where the residual systematics of Sect.~\ref{sec:results-systematics} are most apparent.

After applying the recommended mask of Sect.~\ref{sec:postproc-mask}, the aromatic component carries $72$\% of the combined radiance summed over the footprint, and panel~(b) accordingly follows panel~(a) closely. A second component was selected at $66.9$\% of the fitted pixels; at the remaining $33.1$\% the 3.4--3.5~\micron\ map holds no value (Sect.~\ref{sec:postproc-mask}), which panel~(c) draws as zero and panel~(f) leaves blank. Since the AIC admits the second component only where the data justify its extra parameters, panel~(i) shows where the data support the split, mostly where the band is bright. The choice follows the signal-to-noise ratio of the unscaled uncertainties: a second component is selected at about 30\% of the pixels with $S/N=1$--$10$, 65\% at $10$--$30$ and 97\% at $30$--$100$, and at fixed $S/N$ the fraction barely depends on the coverage. Panels~(d)--(f) contain the statistical uncertainties only, already scaled by the reduced $\chi^{2}$ of each fit (Sect.~\ref{sec:methods-pah-fitting}); the modelling systematics of Sect.~\ref{sec:methods-pah-resolution}, up to 2, 6 and 11\% for the combined, 3.3~\micron\ and 3.4--3.5~\micron\ maps, come on top of them.

\begin{figure}[t]
  \centering
  \includegraphics[width=\columnwidth]{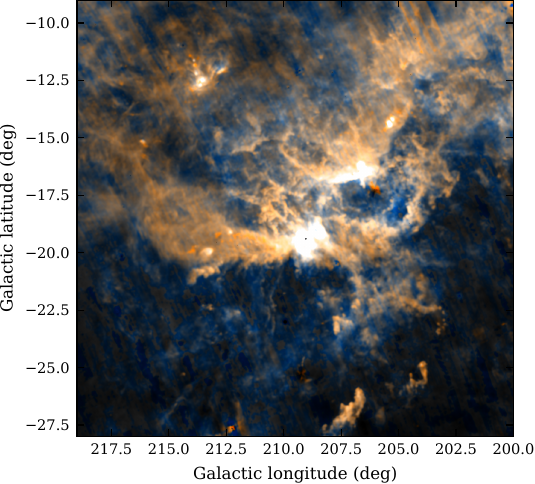}
  \caption{Two-colour view of the aromatic/aliphatic split over Orion A and B ($19\degr$ field, Galactic coordinates), from the destriped maps smoothed to $4\arcmin$. The colour compares the ratio of the two components with its typical value in the field ($0.40$, defined in Sect.~\ref{sec:results-map}): white sightlines have the typical ratio, red ones a relatively stronger 3.4--3.5~\micron\ component and blue ones a relatively stronger 3.3~\micron\ component. Because aliphatic groups are eroded by far-ultraviolet photo-processing, red marks gas more shielded from the ultraviolet field than the field average, as in the molecular clouds, and blue gas that is more strongly irradiated; the colour itself measures only the band ratio. The blue, red and green channels are the 3.3~\micron\ radiance, the 3.4--3.5~\micron\ radiance divided by $0.40$, and their mean, on a common stretch with the colour saturation enhanced, so brightness follows the band radiance. No mask is applied.}
  \label{fig:orion-two-colour}
\end{figure}

As an illustrative example, Fig.~\ref{fig:orion-two-colour} shows how the split varies within a single region, Orion A and B. Relative to the typical ratio of the two components in the field, the median of the 3.4--3.5 to 3.3~\micron\ ratio where both are fitted and the combined radiance is detected at ten sigma or more ($0.40$), the 3.4--3.5~\micron\ component is stronger across the molecular clouds and weaker in the more diffuse gas around them. This is analogous to the variations of the 3.40/3.29~\micron\ band ratio within reflection nebulae that \citet{Joblin1996} attribute to the photochemical erosion of methylated PAHs, and to the higher aliphatic-to-aromatic hydrogen ratio that \citet{Boersma2026} derive with SPHEREx for dense than for diffuse gas across the north-western photodissociation region of the Iris Nebula, both interpreted as the photo-processing of aliphatic groups where the far-ultraviolet field is harsh. The figure is an illustration rather than a measurement: in the faintest parts of the field the residual systematics of the split (Sect.~\ref{sec:results-systematics}) are comparable to its colour contrast.

The absolute uncertainties are largest where the emission is brightest. This follows from the scaling of each uncertainty by the square root of its pixel's reduced $\chi^{2}$ (Sect.~\ref{sec:methods-pah-fitting}), which indirectly absorbs the calibration uncertainty between channels and the modelling uncertainties, both of which grow with the brightness of the sky (see below). The coverage of panel~(g) reflects the depth of QR2, whose holdings extend to about July 2026 (Sect.~\ref{sec:data-survey}): most lines of sight have been visited only a few times, and each visit fills only the part of the 102 channels that the LVF passes at that detector position (Sect.~\ref{sec:methods-healpix-accum}). Coverage is deepest around the two ecliptic poles, which the survey revisits on every orbit and where nearly all 102 channels are valid. Fewer than 20 channels are valid over $0.50$\% of the footprint, the pixels flagged as low coverage in Sect.~\ref{sec:postproc-mask}. The footprint is also incomplete for a reason unrelated to depth: the central longitudes of the Galactic plane carry no usable Level-2 data in this release (Sect.~\ref{sec:methods-healpix-accum}) and appear as unobserved sky.

The reduced $\chi^{2}$ of panel~(h) has a median of $4.0$ over the footprint rather than unity, and it rises with the brightness of the sky. Binned by the signal-to-noise ratio of the unscaled uncertainties, its median stays between 3.5 and 4.3 up to $S/N=100$ and reaches 10 at $S/N=100$--$300$ and 65 at $300$--$1000$. A residual that grows in proportion to the signal is the signature of a relative rather than an additive error, and this rise corresponds to a scatter between channels of about 2\% of the signal (16th--84th percentiles 1.3--3.5\%). Measured directly on the channels of about 5000 pixels at $|b|<5\degr$, the robust scatter of the channel values about the fitted model falls to 2.4\% of the model at a channel signal-to-noise ratio of 100--300 and to 1.6\% at 300--1000, and adding a 2\% term in quadrature to the channel variances lowers the median reduced $\chi^{2}$ of those pixels from 11 to 2.3. That is the level of the channel-to-channel calibration accuracy targeted for SPHEREx, 2\% \citep{Bock2026}; the in-flight calibration is currently reliable at the $\sim$10\% level \citep{SPHERExExpSupp2025}, and we do not include a calibration term in the fit. The spectral model adds a second contribution in H\,{\sc ii} regions, where the hydrogen recombination lines Brackett-$\beta$ at $2.626$~\micron, Pfund-$\epsilon$ at $3.039$~\micron\ and Pfund-$\gamma$ at $3.741$~\micron\ fall in the continuum windows. Brackett-$\beta$, at the blue end of the continuum fit, also bends the baseline there. The reduced $\chi^{2}$ map is released as measured, with no normalisation. The radiance uncertainties, by contrast, have already been multiplied by the square root of each pixel's reduced $\chi^{2}$ (Sect.~\ref{sec:methods-pah-fitting}), so both contributions are included in them. A reduced $\chi^{2}$ above unity therefore does not mean that the released uncertainties are too small, and they should not be rescaled by it again. Values an order of magnitude above the median therefore mark either the brightest sightlines, among them the brightest H\,{\sc ii} regions, or individual failures, and the $2.4$\% of the footprint above $\chi^{2}_{\rm red}=50$ is flagged by the mask. Several such regions have the shape and size of a single $3.5\degr$ Level-2 tile, which points to individual tiles, most likely with a calibration mismatch relative to their neighbours, rather than to anything on the sky.

\subsection{Residual Systematics}
\label{sec:results-systematics}

Two families of stripes affect the band-radiance maps: the degree-scale stripes of Sect.~\ref{sec:postproc-destripe}, which are filtered from the coadded maps, and arcminute lines, which are mostly removed at their source in the spectral fit (Sect.~\ref{sec:methods-pah-wavelengths}). Here we summarise how far each released product can be trusted against them.

In the unfiltered maps the degree-scale stripes are the dominant systematic in faint sky. Measured as the excess variance of latitude over longitude profiles in $3.5\degr$ boxes, their rms in the faint half of the footprint is $9.3\times10^{-11}$~W~m$^{-2}$~sr$^{-1}$ in the combined radiance, comparable to the sky itself, and $5.0\times10^{-11}$~W~m$^{-2}$~sr$^{-1}$ in the aromatic component. Destriping lowers both to $1.6\times10^{-11}$~W~m$^{-2}$~sr$^{-1}$, by 82 and 69\%. The ratio of the variance along the ecliptic parallels to that perpendicular to them, which is 1 for isotropic sky, falls from 18 to 2.1 in the faintest sky, and the correlation with WISE 12~\micron\ (Sect.~\ref{sec:discussion-wise}) rises from $r_{\rm S}=0.67$ to $0.76$ in the fainter half of the compared cells, split at the median 12~\micron\ intensity, while barely changing in the brighter half ($0.952$ and $0.957$), as expected if stripes are removed and sky is not. What remains in the faintest fields is dominated by the outlines of individual tiles whose offsets the filter does not fully level (Fig.~\ref{fig:destripe-m31}b). For the combined and aromatic radiances the filtered maps are therefore the better choice for diffuse emission everywhere.

The arcminute lines follow the LVF curves of each tile (Fig.~\ref{fig:tile-overview}) and change abruptly at tile edges, with a fundamental period of 2.0--$2.4\arcmin$ and its harmonics. In the released maps, fitted with per-sample wavelengths, they are small: in the brightest parts of a $15\degr$ field around $\lambda$~Orionis, structure along the lines has an rms of 0.3\% of the local signal in the combined radiance, 0.2\% in the aromatic component and 1.7\% in the 3.4--3.5~\micron\ component. Because they scale with the signal they are far below the noise in faint sky, and because they are confined to scales of a few arcminutes they average down further when the maps are smoothed to a coarser resolution. For the combined and aromatic radiances they are negligible.

The 3.4--3.5~\micron\ component is less reliable. Its filtered map is the difference of the other two, with a fade-out set by the combined radiance, so its degree-scale stripes are only about halved in faint sky, and reduced by under a third in intermediate sky, and remain comparable to its own signal in the faintest fields. The spectral fit adds a second limitation, independent of scale: it holds the aromatic band rigid, so any residual of the spectral model has one place to go, the aliphatic component, and a pattern with a period of $8$--$15\arcmin$ in the split between the two components, aligned with the arcminute lines along the curves of constant wavelength of each tile, and so broadly with the ecliptic parallels, survives both corrections. Letting the width of the aromatic band vary does not remove it but spreads it into the aromatic component, and the pattern does not follow the spectral coverage of the pixels. Both limitations trace back to how individual tiles enter the fit, and we expect them to be removable by the joint treatment of the tile offsets in the processing planned for a future version. For now we release the 3.4--3.5~\micron\ map as an estimate suited to spatially averaged work, and recommend the combined and aromatic radiances for morphology.

\subsection{Comparison with WISE}
\label{sec:discussion-wise}

The natural external check on a diffuse PAH map is the tracer it is meant to complement. We therefore compare the integrated band radiance against the reprocessed WISE 12~\micron\ map of \citet{MeisnerFinkbeiner2014}, in the $N_{\rm side}=1024$ HEALPix form distributed by LAMBDA\footnote{\url{https://lambda.gsfc.nasa.gov/product/foreground/fg_wise12_micron_dust_map_get.html}}. No comparable all-sky product exists at 3.4~\micron: the AllWISE Atlas and unWISE coadds preserve point sources by construction and carry no zodiacal-light treatment, so a $W1$ comparison would largely measure integrated stellar surface density, which is precisely what Sect.~\ref{sec:methods-starremoval} removes from our maps. The 12~\micron\ map has instead had point sources and their optical artifacts subtracted, the known WISE instrumental artifacts mitigated, and the zodiacal light removed not by subtracting a model, which the authors tried and abandoned, but by fitting a gradient to each exposure against the SFD 100~\micron\ map and replacing the largest angular scales with those of Planck 857~GHz, so the two data sets have been prepared with the same intent. It also carries a physical advantage: $W3$ contains the 11.3~\micron\ aromatic C--H out-of-plane bending mode \citep{Leger1984,Tielens2008}, emitted by larger PAHs than the 3.3~\micron\ stretch \citep{Schutte1993,Draine2007}, so the ratio of the two bands is sensitive to the grain size distribution, though it also depends on the charge state of the emitters and on the hardness of the exciting radiation field.

Before comparison the two maps are brought onto a common resolution and pixelisation: both are smoothed to a common $12\arcmin$ FWHM and then downgraded into $N_{\rm side}=256$ cells. LAMBDA lists a resolution of $12\arcmin$ for the HEALPix version\footnote{\url{https://lambda.gsfc.nasa.gov/product/foreground/fg_wise12_micron_dust_map_info.html}}, but it has not been smoothed: its angular power spectrum follows that of the sky out to its $3.4\arcmin$ pixel scale, so we treat its resolution as set by its pixels. The $13.7\arcmin$ cell is comparable to the common beam, so the compared cells are close to independent.

The pixels that enter are selected with the recommended reliability mask of Sect.~\ref{sec:postproc-mask} rather than with cuts of their own: every native pixel flagged in any of its quality conditions is removed before the smoothing, and a cell is kept only if at least half of it survives. Consistent with Sect.~\ref{sec:methods-pah-lowsnr}, no significance cut is applied and cells of low or negative radiance are retained, since discarding the negative half of a noise-dominated distribution would bias the recovered slope.

\begin{figure}
  \centering
  \includegraphics[width=\columnwidth]{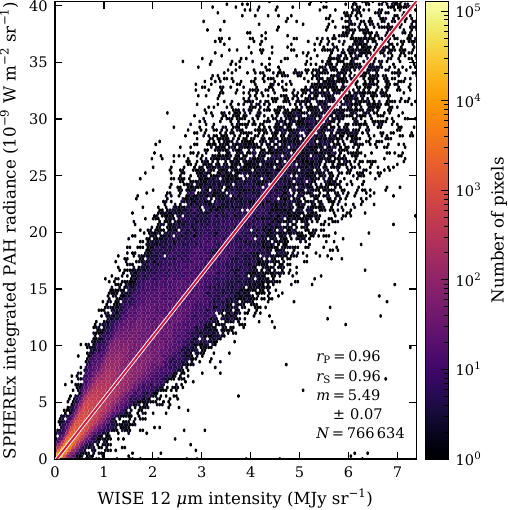}
  \caption{Pixel-by-pixel comparison of the SPHEREx combined PAH radiance, after destriping, against the reprocessed WISE 12~\micron\ intensity of \citet{MeisnerFinkbeiner2014}. Both maps are brought to a common resolution of 12$\arcmin$ and averaged into $N_{\rm side}=256$ cells of $13.7\arcmin$, so that the plotted cells are approximately independent; the colour scale gives the number of cells per bin. Pixels flagged by the recommended reliability mask (Sect.~\ref{sec:postproc-mask}) are excluded before the averaging. The line is a fit of slope $m$, in $10^{-9}$\,W\,m$^{-2}$\,sr$^{-1}$ per MJy\,sr$^{-1}$, to all cells, including the negative ones and the few beyond the plotted range; no significance cut is applied (Sect.~\ref{sec:methods-pah-lowsnr}).}
  \label{fig:tt-wise12}
\end{figure}

Figure~\ref{fig:tt-wise12} shows the result for the destriped combined radiance over the $766\,634$ cells, $40\,214$~deg$^{2}$ or 97\% of the sky, in which at least half of the native pixels are observed and pass the mask, of the 99.4\% covered at present. The two tracers correlate closely, with $r_{\rm S}=0.96$: the spectrally extracted band radiance and an independently processed broadband 12~\micron\ intensity recover the same diffuse structure, which is a mutual confirmation of the two maps over this footprint. The slope, from a Theil--Sen fit that is insensitive to the few extreme cells, is not a calibration, and we do not interpret it as one. The 12~\micron\ band is broad and contains continuum emission alongside the 11.3~\micron\ feature, so a colour-dependent slope, and scatter in excess of the statistical uncertainties, are expected rather than evidence of a defect in either map.

\subsection{Reliability Mask}
\label{sec:postproc-mask}

The integrated radiance is defined at almost every sky pixel the survey has visited, but it is not equally trustworthy everywhere. We therefore release a per-pixel reliability mask alongside the radiance, as a bitmask with one bit for each of the six conditions below, in the order listed (bit 0 to bit 5). It is a mask of known issues, not a detection threshold: it says where a value should not be relied on. The thresholds were chosen by inspecting the flagged pixels against the maps and are fixed for the release.

\begin{itemize}
\item \textit{Ultrabright stars.} The 197 stars of Sect.~\ref{sec:methods-ultrabright} are flagged within the same $r=200$~pixel ($20.5\arcmin$) radius that is masked during the subtraction itself. Their subtraction uses a single stacked model, whereas their wings are shaped by the optics in ways that differ from star to star, so this is where the stellar subtraction is least reliable and where the largest residuals, of either sign, are expected, particularly around the reddest of these stars, such as Betelgeuse.
\item \textit{Residual bright stars.} Stars brighter than $W1=5$, the bright regime of Sect.~\ref{sec:methods-psf-fitting}, are subtracted by the standard branch but may leave a residual, and this is the one condition tested on the map itself rather than predicted. For each star we compare the pixels within $1.5\arcmin$ of its position with the median of a surrounding annulus at $3$--$6\arcmin$, and flag it if either the mean of those pixels departs from the annulus by more than $5\sigma$, or any single one of them by more than $10\sigma$, with $\sigma$ the local noise\footnote{The larger of the propagated uncertainty and the scaled median absolute deviation of the annulus; for the mean it is divided by $\sqrt{N}$, with $N\approx10$ the number of pixels averaged.}. The first test catches a residual spread over the core, the second one too compact to move the mean; both use the absolute departure, so over- and under-subtraction are caught alike. A flagged star is masked within $1.7\arcmin$, the saturated core of Sect.~\ref{sec:methods-psf-fitting}.
\item \textit{Low coverage.} Pixels for which fewer than 20 of the 102 channels carry a valid measurement, so that the spectral fit is poorly constrained regardless of how well the individual channels behaved.
\item \textit{Poor fits.} Pixels whose reduced $\chi^{2}$ exceeds $50$. Every pixel carries a common-mode residual (Sect.~\ref{sec:results-map}) that places the typical value at $\chi^{2}_{\rm red}\approx4$, so the threshold is set an order of magnitude above it. The flag is meant for individual failures, namely a miscalibrated tile, a poorly subtracted source, or a bad channel, but it also sets on the brightest sightlines, where the calibration floor and the recombination lines of Sect.~\ref{sec:results-map} raise $\chi^{2}$; users interested in the brightest H\,{\sc ii} regions can ignore it there, since their uncertainties already include the excess. It also captures the isolated, strongly negative values left where a stellar subtraction failed outright, whose spectra the model cannot fit either.
\item \textit{Unconstrained fits.} Pixels whose fitted radiance has an uncertainty above $3\times10^{-8}$~W\,m$^{-2}$\,sr$^{-1}$ in any of the three maps, some 330 times the median and 14 times the 99th percentile. There the second component runs away, to values of up to $10^{13}$ times the typical radiance with uncertainties larger still, and it takes the combined radiance with it. Almost all of them, 95\%, lie in the inner Galactic plane ($|l|<60\degr$, $|b|<5\degr$), where coverage is thin: they have a median of 20 valid channels, against 57 over the footprint. The condition tests the uncertainty rather than the value, so it introduces no significance cut.
\item \textit{Single-component fits.} Pixels at which the AIC selected $N=1$, so that the 3.4--3.5~\micron\ map holds no value rather than a small one. These are not unreliable pixels; the flag exists so that an absent component is not read as a measured zero.
\end{itemize}

Of the observed footprint, the pixels with at least one valid channel, the five quality conditions flag $0.19$, $0.19$, $0.50$, $2.42$ and $0.20$\% respectively, and their union, the recommended mask, $3.08$\%; the missing aliphatic measurement condition applies to a further $33.0$\%, on which the combined radiance is unaffected.

\begin{figure}[t]
  \centering
  \includegraphics[width=\columnwidth]{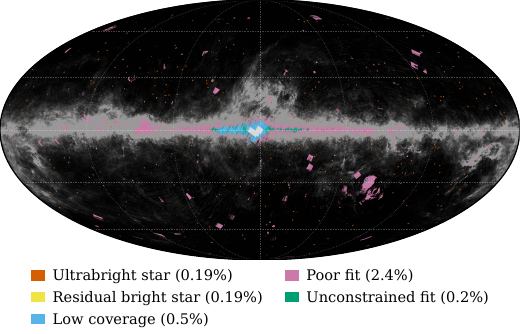}
  \caption{The quality conditions of the reliability mask, in colour (the unconstrained fits, in green, lie along the inner Galactic plane), over the combined radiance in greyscale, in the projection of Fig.~\ref{fig:fullsky-maps}. The mask is drawn to scale: each pixel of the figure, about $4\arcmin$ across, shows the mask value at its centre. The legend gives the fraction of the observed footprint carrying each flag. The single-component flag is not shown.}
  \label{fig:reliability-mask}
\end{figure}

The recommended mask is the union of the first five conditions, shown in Fig.~\ref{fig:reliability-mask}; the sixth concerns only the 3.4--3.5~\micron\ map.

Two points of usage follow, and we state them explicitly because they are easy to miss. Users who smooth the released maps to a coarser resolution should mask first and then interpolate across the masked pixels: a bright residual left inside a mask is spread over the larger beam by the smoothing and contaminates an area far exceeding the mask itself, which is precisely the failure the mask exists to prevent. And the masked sky fraction should be propagated into any power spectrum or correlation analysis rather than treated as zero-signal sky.

\section{Conclusions}
\label{sec:conclusions}

We have presented full-sky maps of the integrated 3.3~\micron\ aromatic band and the 3.4--3.5~\micron\ aliphatic complex from the SPHEREx QR2 Level-2 data of Detector~4, at $2.6\arcmin$ resolution on an $N_{\rm side}=4096$ grid, in which the bands are fitted in a per-pixel spectrum of 102 narrowband channels rather than inferred from broadband photometry. The combined band radiance is the primary product: it changes by at most a few per cent under the modelling choices the data leave open, and it follows the WISE 12~\micron\ map closely ($r_{\rm S}=0.96$) while isolating the band emission itself. The maps are released in Galactic coordinates with their statistical uncertainties, per-pixel fit diagnostics and a bitwise reliability mask, in destriped and unfiltered versions.

The maps are meant as a general resource for the study of the smallest interstellar grains. Normalised by the dust column and by a proxy for the heating starlight, the band isolates environmental changes in the smallest PAHs, as \citet{Li2026} showed for the Magellanic Clouds, where it is suppressed inside the H\,{\sc ii} regions of the LMC; across the Galaxy this follows them from the diffuse medium and molecular clouds to photodissociation regions and, with the hydrogen recombination lines, into the ionised gas in which they are depleted \citep{Murgia2026}. Its ratio to the longer-wavelength bands of larger PAHs constrains their size distribution, and the aromatic--aliphatic split, averaged over regions, their photo-processing (Fig.~\ref{fig:orion-two-colour}). The maps also offer a Galactic reference for the 3.3~\micron\ emission of galaxies, a spectral calibration for broadband PAH proxies, and a template for the Galactic emission near 3~\micron\ wherever it is a foreground. Finally, set against the all-sky spinning-dust maps of \citet{Hoerning2026}, they allow a direct test of whether the smallest PAHs carry the anomalous microwave emission and, if they do, provide a template for it in the separation of the foregrounds to the cosmic microwave background.

Three limitations should be kept in mind. The split between the two components depends on the model, by up to 6\% for the 3.3~\micron\ band and 11\% for the 3.4--3.5~\micron\ complex, which is best used averaged over regions. The degree-scale stripes left by zodiacal-light and calibration residuals are filtered from the coadded maps in faint sky rather than removed at their source, and a weaker residual remains in the faintest fields. And at QR2 depth the per-channel coverage is incomplete, which limits the faintest sightlines.

This is a first release, and the largest gains ahead are in coverage. The accumulator of Sect.~\ref{sec:methods-healpix-accum} is additive and resumable, so further QR2 data refill the same maps at no cost, and the per-channel completeness that limits the faintest sightlines improves with every added pass. The very recently released QR3 should recover the crowded inner Galactic longitudes that QR2 leaves empty, and revises the zodiacal-light model, but the absolute gain and the spectral calibration both changed between the two, so incorporating it means reprocessing rather than refilling. It would not remove the need for the empirical point-spread function of Sect.~\ref{sec:methods-psf-construction} either: the released model is sampled more finely across the focal plane, but its stamp still spans about an arcminute, against the degree-scale wings that dominate the contamination. The residual stripes are the other priority. The arcminute lines are removed at their source, by fitting every sample at its own wavelength, but the degree-scale stripes are only filtered from the coadded maps, and only in faint sky, and the 3.4--3.5~\micron\ component retains both families at a level that restricts it to spatially averaged use. Fitting per-tile offsets and gradients during the processing, as \citet{MeisnerFinkbeiner2014} did per exposure for WISE, or fitting the zodiacal light, would address both at their source, and a spectral treatment of the wavelength spread such as the interpolation of \citet{Cukierman2026} offers a complementary route.

Several smaller improvements are also planned. The uncertainties would benefit from an explicit calibration term, derived empirically from the scatter between channels, in place of the reduced-$\chi^{2}$ scaling, and the spectral model from a treatment of the hydrogen recombination lines that fall in its continuum windows, Brackett-$\beta$ and the Pfund series, which it does not yet include. A per-object model for the ultrabright stars, in place of the single stacked profile of Sect.~\ref{sec:methods-ultrabright}, is what would allow their $20.5\arcmin$ masks to be shrunk. The resolution is set by our pixelisation rather than by the instrument, the native pixel being $6.15\arcsec$, so sub-arcminute maps are possible where coverage supports them. Further ahead, extending the extraction onto Detector~5 ($3.82$--$4.42$~\micron) would anchor the continuum beyond the aliphatic complex and place Brackett-$\alpha$ in the same fit, letting the hydrogen recombination series be subtracted per pixel rather than bounded as a field average, and the aromatic--aliphatic split be set by the data rather than by the priors of Sect.~\ref{sec:methods-pah-resolution}.

\begin{acknowledgements}
We thank Ari Cukierman for useful discussions on the SPHEREx zodiacal-light and PSF models. The author(s) are grateful for the contribution of Dr. \'{A}ngel de Vicente and the IAC High-Performance Computing and hardware facilities to the results of this research.

We also thank Roberta Paladini for helpful discussions, and the IRSA Help Desk for their assistance with access to the SPHEREx QR2 spectral images.

RCA, JARM and RGS acknowledge funding from the Horizon Europe research and innovation program under GA 101135036 (RadioForegroundsPlus) and the Severo Ochoa Centre of Excellence accreditation awarded to the Instituto de Astrofisica de Canarias, grant CEX2025-001609-S. RGS acknowledges funding from  the Spanish MCIN/AEI/10.13039/501100011033, project PID2023-150398NB-I00. JARM acknowledges financial support from the Spanish MCIN/AEI/10.13039/501100011033, project PID2023-151567NB-I00.
CD and SEH acknowledge funding from an STFC (Consolidated Grant ST/P000649/1) and UKSA (LiteBird UK ST/Y005945/1).
GAH acknowledges the funding from the Dean's Doctoral Scholarship by the University of Manchester.
This research is based on observations made with the SPHEREx mission and makes use of data products from the SPHEREx Science Data Center, distributed by the NASA/IPAC Infrared Science Archive (IRSA), which is funded by the National Aeronautics and Space Administration and operated by the California Institute of Technology.
This publication makes use of data products from the Wide-field Infrared Survey Explorer, which is a joint project of the University of California, Los Angeles, and the Jet Propulsion Laboratory/California Institute of Technology, funded by the National Aeronautics and Space Administration.
Some of the results in this paper have been derived using the HEALPix package \citep{Gorski2005} and its Python implementation \texttt{healpy} \citep{Zonca2019}, and make use of \texttt{astropy} \citep{astropy2013,astropy2018,Astropy2022}, \texttt{numpy} \citep{Harris2020}, \texttt{scipy} \citep{Virtanen2020}, \texttt{matplotlib} \citep{Hunter2007} and \texttt{lmfit} \citep{Newville2014}.
\end{acknowledgements}

\bibliographystyle{aa}
\bibliography{paper_refs}

\label{lastpage}

\end{document}